\documentclass[manuscript]{aastex}
\usepackage{graphicx}
\usepackage{subfigure}
\usepackage{rotating} 

\newcommand{\lsim }{{\lower0.8ex\hbox{$\buildrel <\over\sim$}}}
\newcommand{\gsim }{{\lower0.8ex\hbox{$\buildrel >\over\sim$}}}

\def\simge{\mathrel{%
  \rlap{\raise 0.511ex \hbox{$>$}}{\lower 0.511ex \hbox{$\sim$}}}}
\def\simle{\mathrel{
  \rlap{\raise 0.511ex \hbox{$<$}}{\lower 0.511ex \hbox{$\sim$}}}}

\newcommand{\Msun}{\ifmmode {M_{\odot}}\else${M_{\odot}}$\fi}
\newcommand{\Lsun}{\ifmmode {L_{\odot}}\else${L_{\odot}}$\fi}
\newcommand{\Rsun}{\ifmmode {R_{\odot}}\else${R_{\odot}}$\fi}

\usepackage{xargs}
\usepackage{color}

\shorttitle{Multi-frequency, Multi-Epoch Study of Mrk 501 : Hints for a two-component nature of the emission}
\shortauthors{Shukla et al.}
\begin{document}
\title{ Multi-frequency, Multi-Epoch Study of Mrk 501: Hints for a two-component nature of the emission}
\author{A. Shukla}
\affil{Department of High Energy Physics, Tata Institute of Fundamental Research, 
Mumbai 400005, India }
\author{V. R. Chitnis}
\affil{Department of High Energy Physics, Tata Institute of Fundamental Research, 
Mumbai 400005, India }
\author{B. B. Singh}
\affil{Department of High Energy Physics, Tata Institute of Fundamental Research, 
Mumbai 400005, India }
\author{B. S. Acharya}
\affil{Department of High Energy Physics, Tata Institute of Fundamental Research, 
Mumbai 400005, India }
\author{G. C. Anupama}
\affil{Indian Institute of Astrophysics, II Block, Koramangala, 560 034, Bangalore }
\author{P. Bhattacharjee}
\affil{Saha Institute of Nuclear Physics, 1/AF, Bidhannagar, 700 064, Kolkata, India }
\author{R. J. Britto\altaffilmark{\$}}
\affil{Saha Institute of Nuclear Physics, 1/AF, Bidhannagar, 700 064, Kolkata, India }
\author{K.~Mannheim}
\affil{Universit\"{a}t W\"{u}rzburg, 97074 W\"{u}rzburg, Germany}
\author{T. P. Prabhu}
\affil{Indian Institute of Astrophysics, II Block, Koramangala, 560 034, Bangalore }
\author{L. Saha}
\affil{Saha Institute of Nuclear Physics, 1/AF, Bidhannagar, 700 064, Kolkata, India }
\author{P. R. Vishwanath}
\affil{Indian Institute of Astrophysics, II Block, Koramangala, 560 034, Bangalore }

\altaffiltext{\$} {Now~at Department of Physics, University of Johannesburg, PO Box 524, Auckland Park 2006, South Africa}

\begin{abstract}
Since the detection of very high energy (VHE) $\gamma$-rays from Mrk 501, its broad band emission of radiation was mostly and quite effectively modeled using one zone emission scenario. However, broadband spectral and flux variability studies enabled by the multiwavelength campaigns carried out during the recent years have revealed rather complex behavior of Mrk 501. The observed emission from Mrk 501 could be due to a complex superposition of multiple emission zones. Moreover new evidences of detection of very hard intrinsic $\gamma$-ray spectra obtained from {\it Fermi}--LAT observations have challenged the theories about origin of VHE $\gamma$-rays.  Our studies based on {\it Fermi}--LAT data indicate the existence of two separate components in the spectrum, one for low energy $\gamma$-rays and the other for high energy $\gamma$-rays.  Using multiwaveband data from several ground and space based instruments, in addition to HAGAR data, the  spectral energy distribution of Mrk~501 is obtained for various flux states observed during 2011. In the present work, this observed broadband spectral energy distribution is reproduced with a leptonic, multi-zone Synchrotron Self-Compton model. 
\end{abstract}

\keywords{BL Lacertae objects: individual (Mrk~501), Jets}

\maketitle

\section{Introduction}
The BL Lac source Mrk~501 (z=0.034) belongs to a sub-class of active galactic nuclei (AGN) that are known as high-energy peaked blazars (HBL). The broadband emission (radio to $\gamma$-rays) of these objects is dominated by non-thermal radiation which is produced in the innermost part of the jets, that are oriented very close to our line of sight. This broadband emission is strongly Doppler-boosted. Like other TeV blazars, the spectral energy distribution (SED) of Mrk~501 characteristically shows a double-peaked profile. These peaks occur at keV and GeV/TeV energies when the SED is plotted in the $\nu F_{\nu}$ versus $\nu$ representation. The general understanding is that the first hump of the SED is caused by synchrotron radiation from the electron population gyrating in magnetic fields of the jet, but the origin of the GeV/TeV hump is unclear. The composition of these jets is also not known; it is not clear whether they are made of electron-positron plasma or electron-proton plasma. Even though Mrk~501 has been observed over last two decades in the entire electromagnetic spectrum, the existing multifrequency data could not provide explicit answer for the physical mechanisms that are responsible for the production of the GeV/TeV hump. This hump may be produced by interaction of electrons with photons in leptonic models \citep{2004NewAR..48..367K,1993ApJ...416..458D,1996MNRAS.280...67G} or protons with photon fields or magnetic fields in hadronic models \citep{2000NewA....5..377A,2003APh....18..593M,1998Sci...279..684M} or by mixed lepto-hadronic scenario \citep{2012AIPC.1505..635C}. 

The  multifrequency correlations and spectral energy distributions of Mrk~501 were studied extensively in the past by \citep{2000ApJ...538..127S,1998ApJ...492L..17P,1999A&A...347...30V,2000A&A...353...97K,2001ApJ...554..725T,2002A&A...386..833G}, but the nature of this object is still far from being understood. The main reasons for this are the moderate sensitivities of the $\gamma$-ray instruments, and the lack of simultaneous multifrequency data during long periods.

Since the detection of this source at above 500~GeV by the Whipple observatory \citep{1996ApJ...456L..83Q}, it has shown several high states over the entire electromagnetic spectrum and also few orphan TeV flares \citep{2011ApJ...727..129A, 2012A&A...541A..31N}. Mrk~501 is also known for its major, long timescale and short timescale flares in X-rays and VHE $\gamma$-rays \citep{1997ApJ...487L.143C,1998ApJ...492L..17P,2005ApJ...622..160X,2007ApJ...669..862A}. One of its historical outbursts was observed in 1997 when the flux at energies above 1 TeV reached upto 10 Crab \citep{1999A&A...342...69A,1999A&A...349...29A}. Following this outburst, the average flux of VHE $\gamma$-rays dropped to 0.3 Crab during 1998-1999 \citep{2001ApJ...546..898A}.

Rapid, and intra-night variability has been displayed by Mrk~501 over the entire electromagnetic spectrum \citep{2008NewA...13..375G,2007ApJ...669..862A, 2012NewA...17....8G}. Fast variability over time scales of minutes has been detected during TeV orphan flares making the study of this object very interesting \citep{2007ApJ...669..862A}. There have been several mechanisms proposed for producing the observed variability in the jet emission, ranging from plasma mechanisms \citep{1994ApJ...423..172K}, beamed radiation (e.g. \citet{1998A&A...338..399C}), coherent instability in a compact emission region (e.g. \citet{2008MNRAS.384L..19B}), misaligned minijets inside the main jet (e.g. \citet{2010MNRAS.402.1649G}), jet deceleration \citep{2003ApJ...594L..27G, 2007ApJ...671L..29L}, wiggles in an anisotropic electron beam directed along the jet \citep{2009MNRAS.393L..16G}, relativistic plasma blob inside the jet (blob-in-jet model; \citep{2001A&A...367..809K}) and plasma instability such as firehose \citep{2012MNRAS.423.1707S} caused by anisotropic electron beam. 

Some fundamental questions regarding this source such as the content of its jet, location and mechanism of $\gamma$-ray emission and the origin of observed variability are still not answered unambiguously. In an attempt to improve our understanding of the source, we present in this work a detailed study of multiwaveband data taken during 2011 January to 2012 March using ground and space based instruments.
We have fitted multiwaveband SED with multi-zone SSC model and discussed constraints on the physical parameters of Mrk~501.

In this paper, we study the multiwavelength-multi-epoch behavior of Mrk~501 during the year 2011.
Five different flux states at different epochs along with a quiescent state SED observed by \citet{2011ApJ...727..129A} are modeled with a two zone SSC scenario and compared. Multiwavelength observations and analysis is presented in $\S$2, and results are discussed in $\S$3. A description of our new two zone model and modeling of six different flux states are described in $\S$4. Finally we discuss implications of our two zone model on the blazar parameters considering Mrk~501 as an example in $\S$5 . 

\begin{table*}
\centering
\small
\caption{HAGAR observations of Mrk~501 in 2010 and 2011}

\begin{tabular}{ccccc}
\hline
\hline
Epoch & Total duration & Excess number of  & Mean $\gamma$-ray rate & Significance \\
&   (min)  (Pairs)     & ON source events       & (/min) & $\sigma$\\
\hline
2010 March 22-2010 May 20     &  400.2 (10) &  1577.0  $\pm$ 511.9   &  3.9  $\pm$ 1.3  &   3.1  \\
2011 March 31-2011 April 10   &  279.5 (7)  &  989.2   $\pm$ 399.6   &  3.5  $\pm$ 1.4  &   2.5  \\
2011 April 28-2011 May 10     &  348.6 (9)  &  2308.1  $\pm$ 454.9   &  6.6  $\pm$ 1.3  &   5.1  \\
2011 May 26-2011 June 03     &  212.6 (6)  &  967.1   $\pm$ 369.2   &  4.6  $\pm$ 1.7  &   2.6   \\
\hline
\end{tabular}
\label{mrk501_res}
\end{table*}

\section{Multiwavelength observations and analysis}
VHE $\gamma$-rays observations were made using High Altitude GAmma Ray (HAGAR) telescope array, at Hanle, India. In addition, archival data from Large Area Telescope (LAT) onboard {\it Fermi}, Proportional Counters Array (PCA) and All Sky Monitor (ASM) onboard {\it RXTE}, X-ray Telescope (XRT), Ultra-Violet/Optical Telescope (UVOT), and Burst Alert Telescope (BAT) onboard {\it Swift}, SPOL and Owens Valley Radio Observatory (OVRO) were analyzed to obtain light curves and energy spectra. 

\subsection{Optical and radio data}
The optical and radio data made available from {\it Fermi} multiwavelength support program are used. The optical observations were made by SPOL team using the SPOL CCD Imaging/Spectropolarimeter at Steward Observatory \citep{2009arXiv0912.3621S}. The optical V-band photometric and polarimetric fluxes are made publicly available by them on websites\footnote{http://james.as.arizona.edu/\~psmith/Fermi/} and this data is used to obtain light curves and SEDs.

The 15 GHz radio observations were made by using a 40 meter single-dish telescope at Owens Valley Radio Observatory (OVRO).
The radio fluxes are also made publicly available by OVRO collaboration on websites\footnote{http://www.astro.caltech.edu/ovroblazars/} and they are used to obtain light curve at radio wavelengths. Details of the analysis method are described in (\cite{2011ApJS..194...29R}). In addition to this, we have also plotted radio data from Very Long Baseline Array (VLBA) at frequencies 5 GHz, 43 GHz and Submillimeter Array (SMA) at 230 GHz, obtained from \citet{2011ApJ...727..129A} on each SED for the reference, as we do not have any radio observations of the core during 2011-2012.

\subsection{{\it RXTE} and {\it Swift}}
The PCA \citep{1993A&AS...97..355B} onboard {\it RXTE} is an array of five identical xenon-filled proportional 
counter units (PCUs). The PCUs cover energy range from 2\,--\,60 keV with a total collecting area of 6500 cm$^{2}$. We have analyzed standard 2 PCA data that have a time resolution of 16 seconds with energy information in 
128 channels. Data analysis was performed using HEASOFT (version 6.10). Data from PCA were analyzed to obtain the X-ray energy spectrum and light curve. For each of the observations data were filtered using the standard procedure provided in the {\it RXTE} Cook Book. The background models were generated with the tool ``pcabackest'', based on {\it RXTE} GOF calibration files for a 'faint' source (less than 40 ct/sec/PCU).

The XRT onboard {\it Swift} uses a grazing incidence Wolter I telescope to focus X-rays onto a CCD \citep{2005SSRv..120..165B}. The instrument has an effective area of 110 cm$^{2}$, 23.6 arcmin FOV, 15 arcsec resolution (half-power diameter), and covers an energy range of 0.2\,--\,10 keV. The windowed timing (WT) mode data were used to obtain the spectrum from {\it Swift}-XRT. Source photons were extracted using a box region with the length of 40 pixels and width about 20 pixels. Events with grades  0\,--\,2 were selected. The spectral data were rebinned by GRPPHA 3.0.0 with minimum 20 photons per bin. Standard auxiliary response files and response matrices were used.

A combined spectral fit was obtained for PCA and XRT data by normalizing the PCA spectrum with XRT spectrum. The PCA and XRT spectra in the energy range of 0.3\,--\,30 keV were fitted by using XSPEC with a cutoff powerlaw with line-of-sight absorption. The line-of-sight absorption was fixed to a neutral hydrogen column density of 1.56$\times$10$^{20}$ cm$^{-2}$ \citep{2005A&A...440..775K}.

{\it Swift}-XRT light curves are obtained from {\it Fermi} multiwavelength support program websites\footnote{http://www.swift.psu.edu/monitoring/} and used in this study. 

The ``Dwell'' data from {\it RXTE}-ASM were obtained from the ASM website\footnote{http://xte.mit.edu/} and were analyzed
with the method discussed in \citet{2009ApJ...698.1207C}. A daily average flux between 15\,--\,50 keV from {\it Swift}-BAT was
 obtained from BAT website\footnote{http://heasarc.nasa.gov/docs/swift/results/transients/}, a detailed analysis procedure can be found in \citet{2013ApJS..209...14K}. 

The {\it Swift}-UVOT \citep{2005SSRv..120...95R} data were used to obtain fluxes in UVW1, UVM2 and UVW2 filters for different epochs. The snapshots of every individual observation were integrated with {\it uvotimsum} task and then analyzed with the {\it uvotsource} task. A source region of 10$^{\prime\prime}$ radius was selected around the source, while the background was extracted from a circular region of 1$^{\prime}$ which is centered in a source-free region. The flux obtained was corrected for Galactic extinction of E$_{(B-V)}$=0.02 mag as given by \citet{1998ApJ...500..525S} in each spectral band.
\begin{figure}
\centering
\includegraphics[width=9cm]{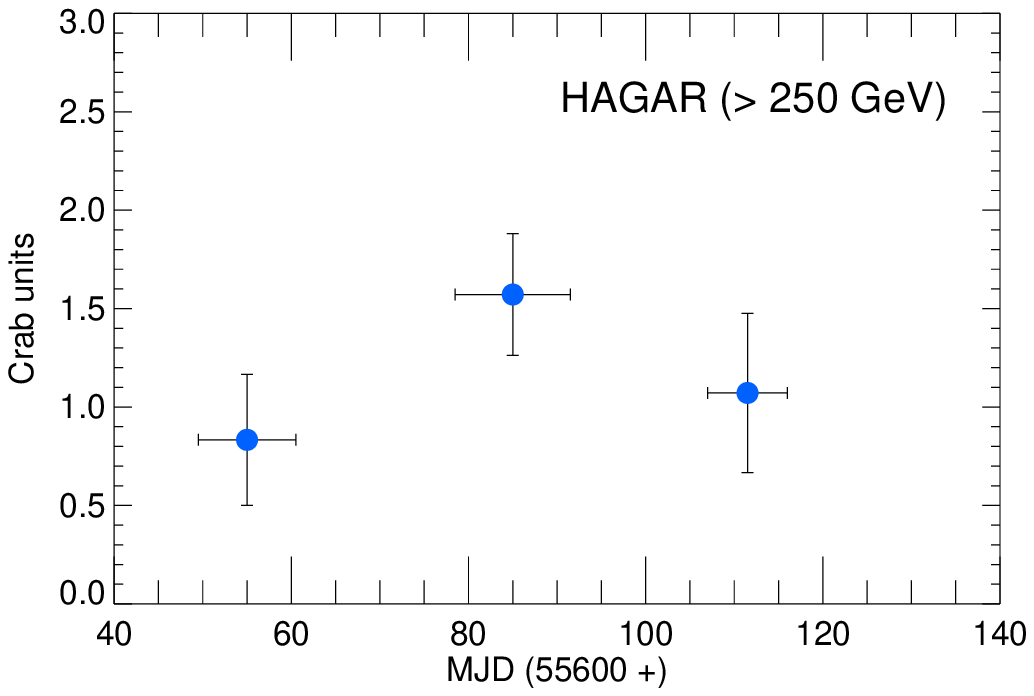}
\caption[]{HAGAR light curve of Mrk~501 during 2011}
\label{hagar_lc}
\end{figure}

\subsection{{\it Fermi}-LAT}
The  {\it Fermi}-LAT is a pair production telescope \citep{2009ApJ...697.1071A} onboard the {\it Fermi} spacecraft. LAT 
covers the energy range from 20 MeV\--\,300 GeV with a field of view $\ge$ 2.5 sr. The {\it Fermi}-LAT $\gamma$-ray data of Mrk~501 over the period of 450 days (MJD: 55560\--\,56010) were obtained from website\footnote{http://fermi.gsfc.nasa.gov/}. Data above 200 MeV were analyzed using the standard analysis procedure {\it(ScienceTools-v9r31p1)} provided by the {\it Fermi}-LAT collaboration.

A circular region of 10$^\circ$ radius ``region of interest (ROI)'' was chosen around Mrk~501 for event reconstruction from the so-called ``diffuse'' event class data which has the maximum probability of being the source photons. Events having a zenith angle 
$<$100$^\circ$ are only retained to avoid the background from Earth's albedo. The spectral analysis of the resulting data set was carried out by 
including galactic diffuse emission component model (gal$\_$2yearp7v6$\_$v0.fits) and an isotropic background component model 
(iso$\_$p7v6source.txt) with post-launch instrumental response function P7SOURCE$\_$V6, using unbinned maximum likelihood 
analysis \citep{1979ApJ...228..939C,1996ApJ...461..396M}. A power law was used to model the source energy spectrum 
above 200 MeV, with integral flux and photon index as free parameters. The flux, and spectrum were 
determined by using unbinned GTLIKE algorithm.
\subsection{HAGAR}
The High Altitude GAmma Ray (HAGAR) telescope array is an Atmospheric Cherenkov Telescope array used for detecting VHE $\gamma$-rays from the celestial sources. This array uses the wavefront sampling technique, and is located at the Indian Astronomical Observatory (IAO), 
Hanle (32$^\circ$ 46$^{\prime}$ 46$^{\prime\prime}$ N, 78$^\circ$ 58$^{\prime}$ 35$^{\prime\prime}$ E), in the Ladakh region of  India, at an altitude of 4270~m. HAGAR consists of an array of seven telescopes arranged in the form of a hexagon, with one telescope at the center. Each telescope is separated by 50~m distance from its neighboring telescope. These telescopes use alt-azimuth mounting system \citep{2013ExA....35..489G}. All seven telescopes have seven para-axially mounted front coated parabolic mirrors of diameter 0.9~m in each, with a UV-sensitive photo-tube at the focus of individual mirrors. These parabolic mirrors have f/D ratio of 1 and they were fabricated by using 10 mm thick float glass sheets. FOV of HAGAR telescope is 3$^{\circ}$ FWHM. The photomultiplier tubes (PMT) which are mounted at the focus of these mirrors are manufactured by Photonis (XP2268B) and have a peak quantum efficiency of 24$\%$ at 400~nm. The high voltages fed to these PMT are monitored and controlled by C.A.E.N controller module (SY1527). In addition to recording signals from individual PMTs, the signals from the 7 PMTs of a telescope are linearly added to form a telescope pulse which is also recorded. The presence of any 4 telescope pulses above a preset threshold value and within a window of about 150 ns forms the trigger for initiating data acquisition. The typical trigger rate was about 12 Hz.

The CAMAC based Data Acquisition (DAQ) system is used in HAGAR. Relative arrival time of the Cherenkov shower front at each
mirror is recorded for each event, as measured by TDCs with a resolution of 250 ps. The Cherenkov photon density at each telescope is measured by the total charge present in PMT pulses using 12 bit QDCs and a Real Time Clock (RTC) module synchronized with GPS is used to record the absolute arrival time of these events accurate upto $\mu$s. In addition to this, a parallel DAQ using commercial waveform digitizers with a sampling rate of 1 GS/s (ACQIRIS make model DC271A) is also used to record telescope pulses.

The energy threshold of the HAGAR telescope array is estimated to be 208~GeV for vertically incident $\gamma$-ray showers for a $\ge$ four-fold trigger condition, for which the corresponding collection area is 3.44$\times$10$^{8}$ cm$^{2}$. The corresponding sensitivity is such that HAGAR will detect a Crab-nebula-like source at a significance level of 5$\sigma$ in 17 hours of observation \citep{2013APh....42...33S}, with no additional criteria for the rejection of background cosmic ray events.

Observations of Mrk~501 were made during March-May 2010 and March-June 2011 on moon-less, clear nights using HAGAR telescope. Observation details are provided in Table \ref{mrk501_res}. The observations were carried out by tracking the source or background region with all seven telescopes. Each source (ON run) was followed (or preceded) by a background (OFF run) with the same exposure time (typically 40 minutes) covering the same zenith angle range as that of the source to ensure that observations were carried out at almost the same energy threshold. 
Selection criteria were applied to identify good quality data. Data were analyzed according to the procedure discussed in  \citet{2012A&A...541A.140S}. In this procedure , Cherenkov shower front is approximated with plane front and space angle i.e. angle between shower axis and pointing direction of telescope is estimated. $\gamma$-ray signal is estimated comparing space angle distribution from ON-OFF pair. Only events with signals in at least five telescopes ($\ge$ five-fold) were analyzed to reduce systematic errors, which corresponds to an energy threshold of 250 GeV for $\gamma$-rays.

\subsection{ARGO-YBJ}
The ARGO-YBJ  experiment situated at the Yangbajing Laboratory, Tibet at  4300~m a.s.l. was designed to study cosmic $\gamma$-radiation, at an energy threshold of $\sim$ 100 GeV, by means of the detection of small size air showers.  ARGO-YBJ consists of a single layer of Resistive Plate Counters (RPC) detectors covering an area of $\sim$ 6700 m$^{2}$ to detect air showers. We have used published TeV $\gamma$-ray data of Mrk~501 from ARGO-YBJ, collected during 2011 October 17 to November 22, \citep{2012ApJ...758....2B}.
\begin{table*}
\centering
\small
\caption{Time periods used to obtain SEDs}
\begin{tabular}{ccccc}
\hline
\hline
Epoch & MJD & Dates & Days \\
\hline
S1       &  55651 - 55661  &       2011 March 31 - 2011 April 10            & 10     \\
S2       &  55679 - 55692  &       2011 April 28 - 2011 May 11              & 13       \\
S3       &  55707 - 55716  &       2011 May 26 -   2011 June 03               & 9     \\
S4       &  55860 - 55890  &       2011 October 26 - 2011 November 25      & 30        \\
S5       &  55919 - 55934  &       2011 December 24  - 2012 January 08        & 15       \\

\hline
\end{tabular}
\label{SED}
\end{table*}

\begin{figure*}
\centering
\includegraphics[width=14cm]{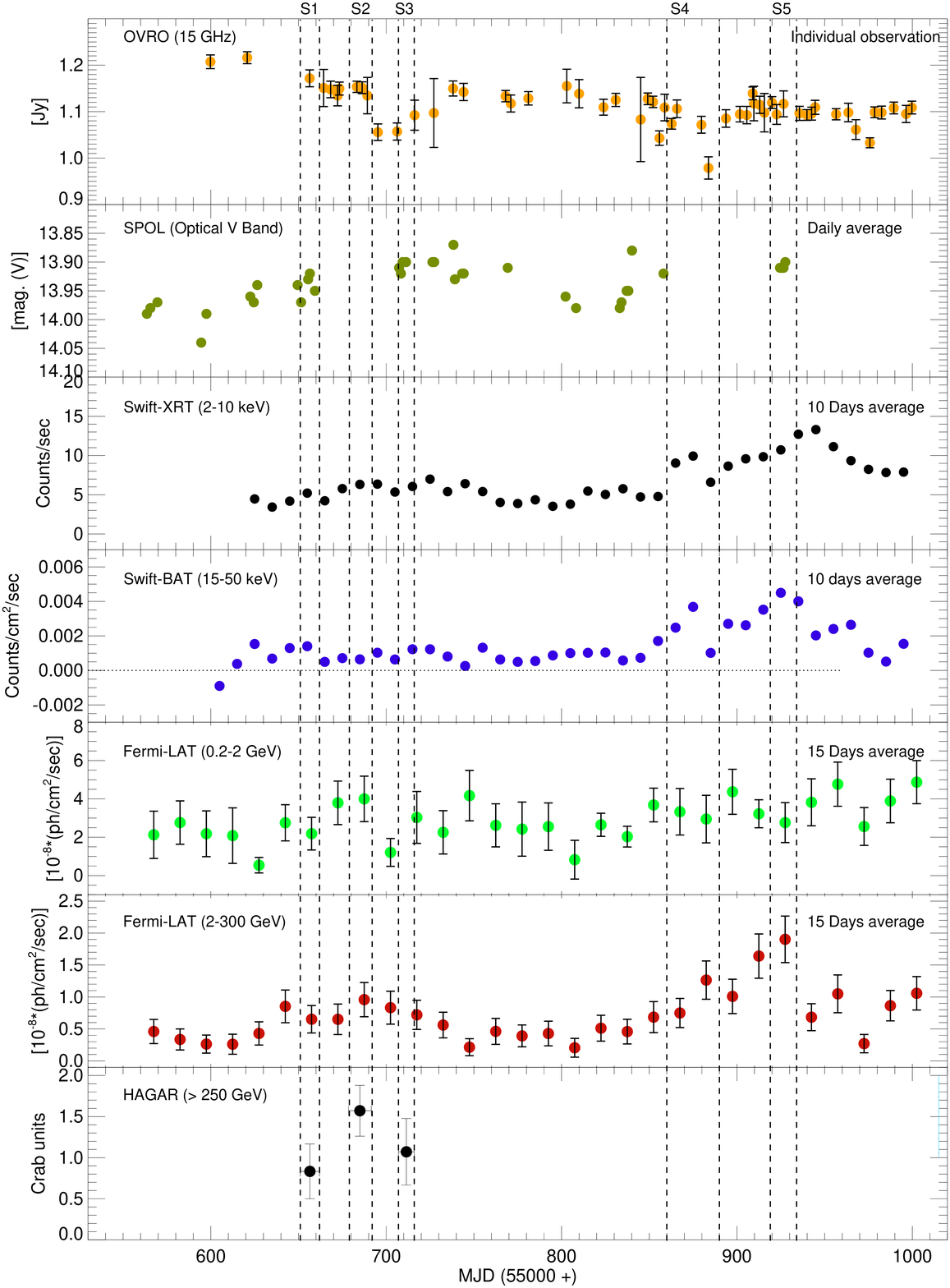}
\caption[]{Multiwavelength light curves of Mrk~501 during 2011-2012. The bin size (in days) used for averaging the flux at different energy bands are mentioned at the right corner. The bin size for HAGAR observation are marked using x-error bars. The vertical dashed lines are representing the periods of moderate and high flux states for which SEDs are obtained, details of these states are provided in Table \ref{SED}.}
\label{MW_ALL}
\end{figure*}

\section{Results}
Mrk~501 was observed during year 2010 and 2011 when it was in a moderate state of activity using HAGAR telescope and VHE $\gamma$-rays were detected from it. The source was detected with a 5$\sigma$ significance during 2011 May (MJD: 55679\,--\,55692) when average flux reached a peak flux of $\sim$1.5 Crab units (1 Crab unit=4.2 counts/min for at least five telescopes triggering). The average integral flux in this observation period above 250~GeV is found to be 4.04$\times$10$^{-10}$ ph/cm$^{2}$/sec. The light curve based on HAGAR observations during 2011 March\,--\,June is shown in Figure \ref{hagar_lc}. 
The TeV $\gamma$-rays from Mrk~501 are detected at a total significance of 6.7$\sigma$ over the observation period of two years. 
\begin{figure}
\centering
\includegraphics[width=10cm]{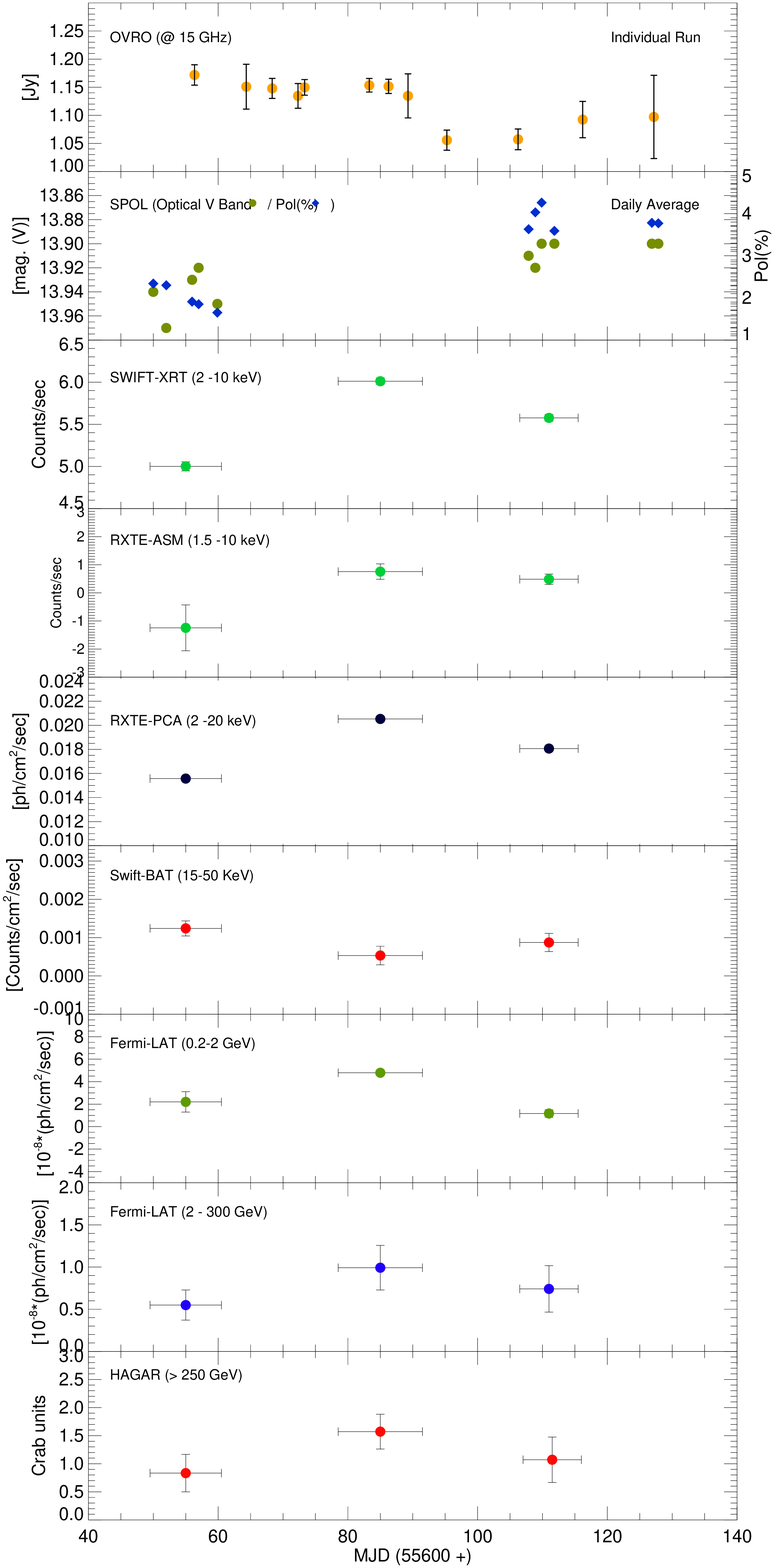}
\caption[]{Multiwavelength light curve of Mrk~501 during HAGAR observations}
\label{Mw_lc_a}
\end{figure}
The source had brightened up moderately during 2011 over the entire electromagnetic spectrum. The multiwavelength light curves, from radio to $\gamma$-rays are used to understand its flux levels and variability. The multiwaveband quasi-simultaneous light curve of Mrk~501 during 2011-2012 is plotted in Figure \ref{MW_ALL}. A few moderate and high states are identified during this period to study the spectral variation with the activity in X-ray and $\gamma$-ray bands. SEDs were obtained for five such states by fitting multiwavelength data with SSC model. Details of these states are provided in Table \ref{SED}. 

\begin{figure}
\centering
\includegraphics[width=8cm]{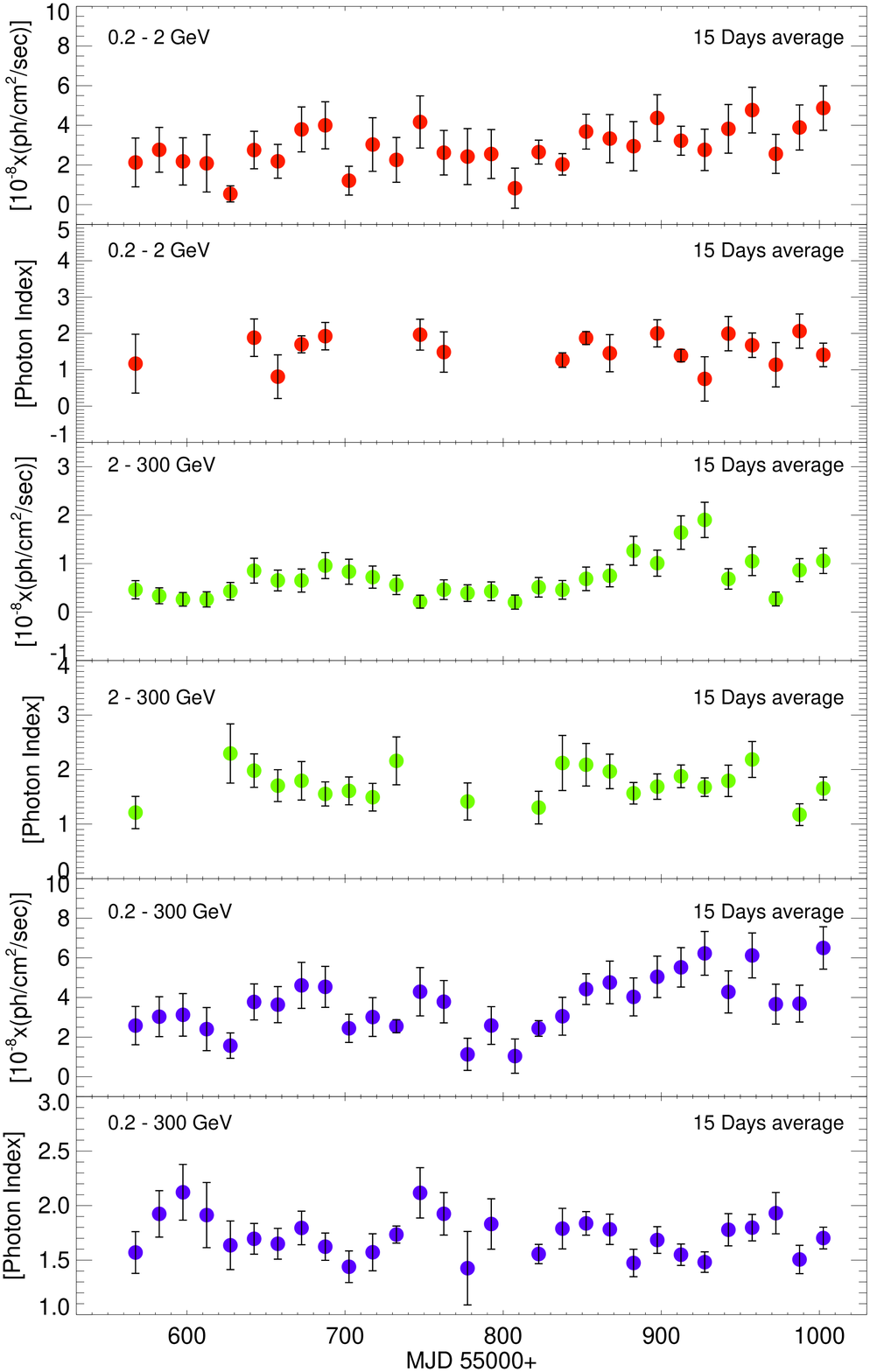}
\caption[]{High energy $\gamma$-ray light curve of Mrk~501 and photon index from {\it Fermi}-LAT during 2011-2012
\label{Fermi_lat}
}
\end{figure}

\subsection{Flux and spectral variation during 2011-2012}
Multiwavelength flux and spectral variability of Mrk~501 as measured and reported by several ground and space based instruments during the 2011-2012 are presented in this section. The multiwaveband quasi-simultaneous light curve of Mrk~501 during 2010 December 30 to 2012 March 24 is shown in Figure \ref{MW_ALL}.
The panels, in a descending order, correspond to data from OVRO (15GHz), SPOL (Optical V band), {\it Swift}\,--\,XRT (2\,--\,10 keV), {\it Swift}\,--\,BAT (15\,--\,50 keV), {\it Fermi}-LAT (0.2\,--\,2 GeV), and {\it Fermi}-LAT (2\,--\,300 GeV). The bottom panel corresponds to HAGAR data above 250 GeV.

Mrk~501 was found to be variable in all the wavebands during the time span of 2011 January to 2012 March with a few active states during this period. A clear variation of flux over a period of an year is observed in the radio, optical, X-rays, and $\gamma$-rays. We found that the source was brightest in X-rays and $\gamma$-rays at the end of the year when it showed couple of flares in X-ray and $\gamma$-rays. During this period of active states a few X-ray flares were observed by {\it Swift}-BAT (15-50 keV). Source showed flaring behaviour during 2011 October 26\,--\,2011 November 25 and the peak flux in this duration was observed on MJD:55873, however peak flux observed by {\it Swift}-XRT was on MJD: 55931.09. The X-ray ({\it Swift}-BAT) light curve shows mild correlation with high energy $\gamma$-ray light curve (2-300 GeV). This correlation appears to be stronger in high state of the source. A similar trend was reported earlier also by \citet{2006ApJ...646...61G}. Fluxes in different wavebands during HAGAR observation periods are also plotted, see Figure \ref{Mw_lc_a}. Source was bright during month of 2011 May in entire electromagnetic spectrum. 

The {\it Fermi}-LAT light curves at different energies (0.2-300 GeV, 0.2-2 GeV, and 2-300 GeV) during the period from 2010 December 30 to 2012 March 24 are plotted with a bin size of 15 days in Figure \ref{Fermi_lat}. The top two panels of this figure correspond to flux and photon index of low $\gamma$-ray energy band (0.2-2 GeV), next two panels correspond to flux and photon index of high $\gamma$-ray energy band (2-300 GeV). The bottom two panels of this figure correspond to flux and photon index of full $\gamma$-ray energy band (0.2-300 GeV). Photon indices are only plotted for those bins that have more than 5$\sigma$ detection.
The LAT observed flux and spectral variation in low energy $\gamma$-ray band (0.2-2 GeV) is found to be different compared to high energy band (2-300 GeV).

Studies of spectral properties of Mrk~501 in X-ray and $\gamma$-ray bands show significant spectral variability during the year 2011 in both the bands. Comparison of spectral variation with the activity of the source was carried out for the period (MJD: 55651-55934) using {\it RXTE}-PCA, {\it Swift}-XRT and {\it Fermi}-LAT instruments. X-ray and $\gamma$-ray spectral indices were obtained for several flux states and given in Table \ref{FS}. This table contains time interval, $\gamma$-ray flux above 0.2 GeV, $\gamma$-ray photon index, and values of likelihood test statistics (TS) as obtained from {\it Fermi}-LAT analysis for which spectra are made and last column contains X-ray indices as obtained from combined fit of {\it Swift}-XRT and RXTE-PCA data.

The {\it Fermi}-LAT collaboration had reported the value of photon index to be 1.78 based on the first 480 days of their observations of Mrk~501. 
This refers to an average spectrum mostly during the quiescent state. They detected remarkable spectral variability where the observed spectral index ranges from hardest value of 1.52$\pm$0.14 to the softest 2.51$\pm$0.20 \citep{2011ApJ...727..129A}. This change in the spectrum was not found to be correlated with the measured flux variations above 0.3 GeV. We also have detected large spectral variability during year 2011 in our study. We found much harder spectra than \citep{2011ApJ...727..129A} during our study, hardest spectrum was detected during S3 with the value of 1.27$\pm$0.21 in energy band of 0.2-300 GeV, (for details see Table \ref{FS}). Using 15 days bins, we have also detected very hard spectra of indcies $\sim$ 1.2 during our 2011-2012 analysis, at higher energy $\gamma$-rays band, (see fourth panel Figure \ref{Fermi_lat}). The spectral variability is also seen in X-ray band during 2011. 

\begin{figure}
\centering
\includegraphics[width=7cm]{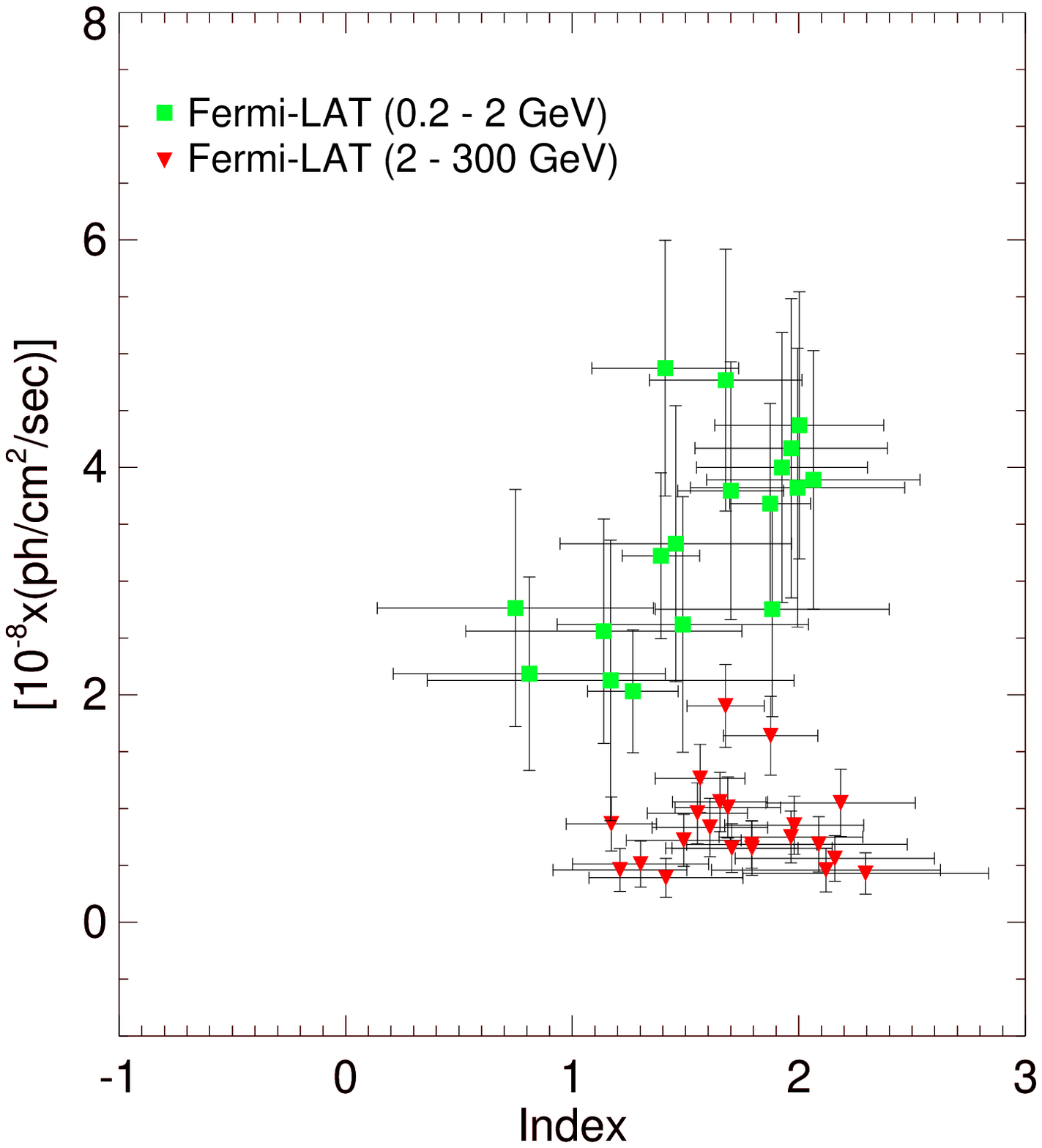}
\caption[]{Cross plot: Flux vs Photon Index during 2011 with 15 days bin 
\label{cross}
}
\end{figure}

The cross plot between $\gamma$-ray flux and photon index, shown in Figure \ref{cross} clearly shows two populations, one for low energy bins of 0.2-2 GeV and second for 2-300 GeV. The photon index increases with the increase in flux for lower energies (plotted in green square), but the cross plot shows scatter in case of high energy $\gamma$-rays and no significant trend is visible (plotted in red downward triangles). This property indicates that low energy $\gamma$-rays may be produced in a different emission zone having slightly different electron energy
distribution and magnetic field than higher energy VHE $\gamma$-rays in Mrk~501.

In addition to the results discussed in this paper, the recent {\it Fermi}-LAT observations of Mrk~501 challenge our present understanding about this source \citep{2012A&A...541A..31N}. New evidence for the presence of very hard intrinsic $\gamma$-ray spectra obtained from {\it Fermi}-LAT observations has challenged the theories of origin of VHE $\gamma$-rays. Several very interesting and viable explanations for the observed hard spectra have been proposed in the recent years. A hard $\gamma$-ray spectrum could be obtained much more easily in hadronic scenario as discussed in proton synchrotron model \citep{2000NewA....5..377A}. Whereas, achieving hard spectrum from leptonic models is more demanding. Some viable scenarios, such as, relativistic Maxwellian-type electron energy distributions which are formed by a stochastic acceleration process as cause of hard spectra  \citep{2011ApJ...740...64L} or the hard spectra produced by an electromagnetic cascade initiated by very-high-energy $\gamma$-rays in the intergalactic medium \citep{2012A&A...541A..31N} are available in the literature.

\section{Spectral Energy Distribution}
\begin{figure}
\centering
\includegraphics[width=9cm]{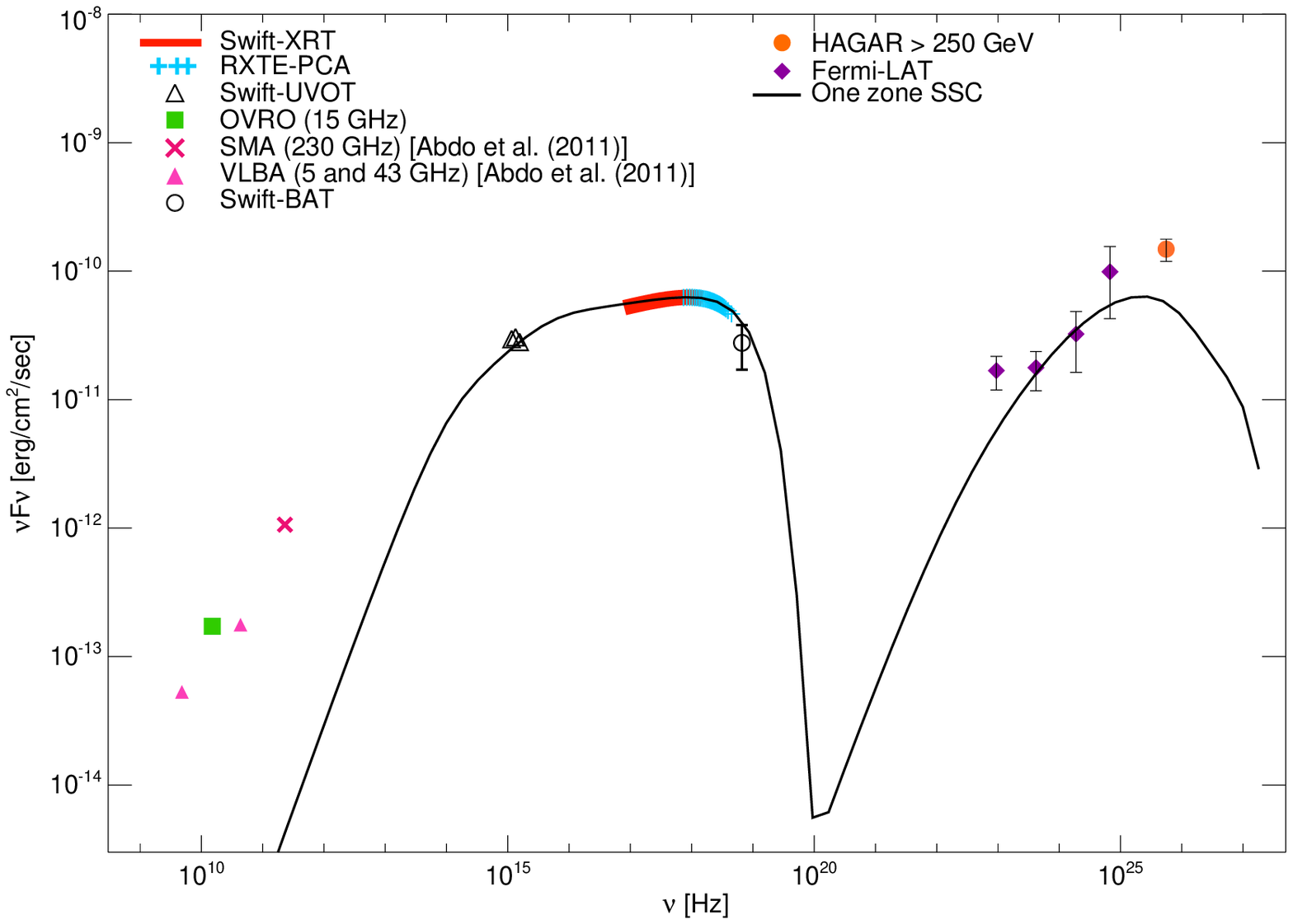}
\caption[]{One Zone SED of Mrk 501 during 2011 April-May [S2]}
\label{one_zone}
\end{figure}
Since its detection at VHE $\gamma$-rays, the broad-band emission from Mrk~501 has been explained mostly using one zone models in literature so far, which was quite effective to explain the observed data from radio to $\gamma$-rays \citep{2000ApJ...538..127S, 2000A&A...353...97K}. The main reason of the apparent success of one zone model was the lack of the simultaneous multiwaveband data and absence of data at low $\gamma$-ray energies.
The broad-band spectral and flux variability studies enabled by the multiwavelength campaigns carried out during the recent years revealed rather complex behavior of Mrk~501. The modeling of Mrk~501 SED assuming a homogeneous single emission zone is a very simplified situation. The broad band emission from blazar may be produced in an inhomogeneous region. The observed emission from Mrk~501 could be due to a complex superposition of multiple emission zones. The study carried out by \citet{2012A&A...541A..31N} and the one presented here indicates the existence of two separate components in the spectrum, one for low energy $\gamma$-rays and one for high energy $\gamma$-rays. For example, emission from single zone is not enough to explain broad band emission of Mrk~501 as shown in Figure \ref{one_zone}, where Mrk~501 SED is modeled using single zone. 

We have considered two zone scenario to explain the broad band SED of Mrk~501. We assume that the observed broad-band SED is sum of two components (two zones) which are radiating simultaneously and boosted with almost same Doppler factor. These two zones are having comoving radii R$_{in}$ and R$_{out}$, and travel with a bulk Lorentz factor $\Gamma$ towards the observer. The emission zones are filled with randomly oriented uniform magnetic fields B$_{in}$ and B$_{out}$ and isotropic population of non-thermal electrons. The energy spectra of the injected electrons in the jet frame are described by broken power laws with low-energy (E$_{min}$ to E$_{b}$) and high-energy (E$_{b}$ to E$_{max}$) components with indices of p1 and p2. The outer zone is responsible for quiescent state flux and other compact inner zone which is close to the black hole is responsible for flaring activity in the jet. The broad band emission from blazar zone is sum of flux of quiescent component and active component.

The radius of the emission zone is constrained by the variability time scales. The comoving radius of the emission zone is defined as 
\begin{equation}
 R\sim{c\delta t_{var}}/{(1+z)}.
\end{equation}
The values of t$_{var}$ for both zones are provided in Table \ref{SED_mrk501}. These values are consistent with the flux variability observed in Mrk~501 during our study period and available in literature (\citep{2011ApJ...729....2A} and references there in).

\begin{figure}
\centering
\includegraphics[width=9cm]{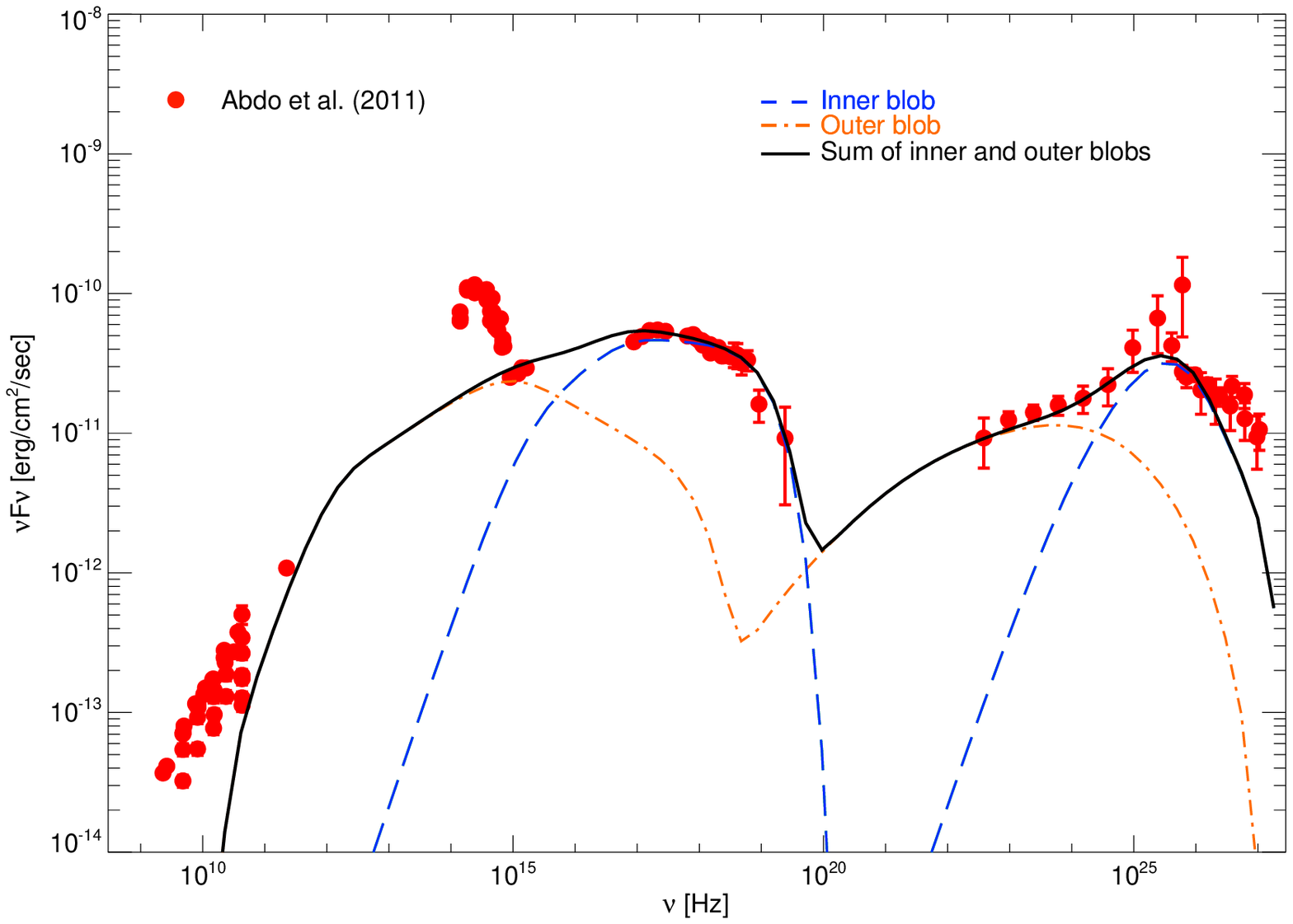}
\caption[]{Two zone Spectral energy distribution of Mrk~501, as observed and presented in MAGIC SED. The solid black line show the best fits two zone model to the data, with the best-fit parameters listed in Table \ref{SED_mrk501}. }
\label{SED_M}
\end{figure}

\begin{figure}
\centering
\includegraphics[width=9cm]{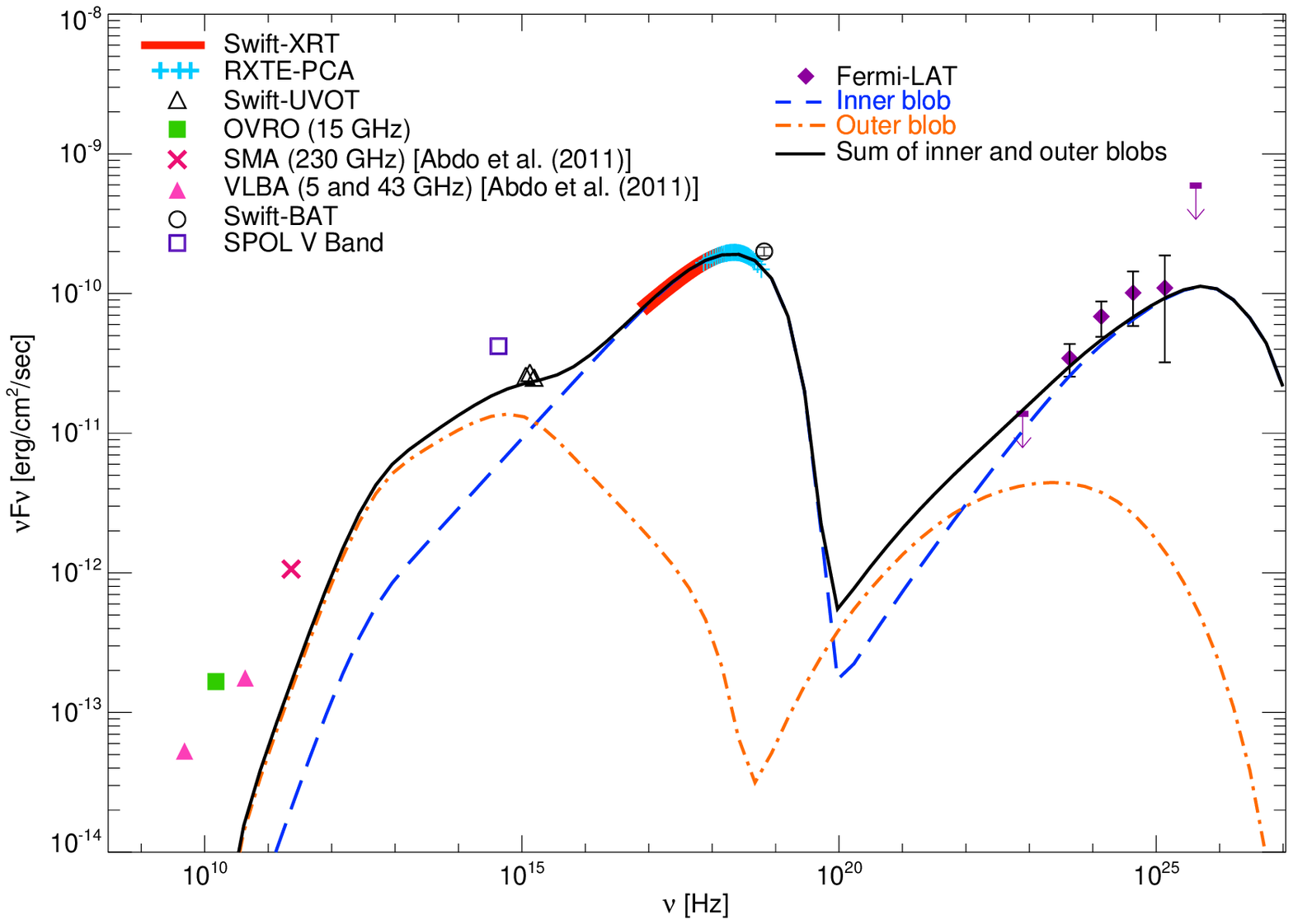}
\caption[]{Two zone SED of Mrk 501 during 2011 December [S5]}
\label{SED_D}
\end{figure}

Five different flux states have been identified during 2011, when the source was in moderate and high state of activity (see Figure \ref{MW_ALL} and Table \ref{SED}). An average TeV spectrum as observed by ARGO-YBJ during October 17 to November 22 is used to model S4 state. The SED obtained as a result of the 4.5-month-long multifrequency campaign (2009 March 15\--\, 2009 August 01) organized by {\it Fermi} and MAGIC collaboration (here after MAGIC SED) \citep{2011ApJ...727..129A} is also modeled with two zone model for comparison. Multiwaveband data for all six SEDs (S1 to S5 and MAGIC SED) are fitted with two zone SSC model and satisfactory fits are obtained, (see Figures \ref{SED_M}\--\,\ref{SED_D}). One more point to be noted is that BAT flux covering the energy range of 15-50 keV during S1 state is somewhat higher than the fitted model, whereas fitted model agrees well with the data from Swift-XRT and RXTE-PCA together covering the energy range of 0.3-30 keV. For other states, BAT flux seems to roughly agree with the fitted model and measurements from lower energy X-rays'.

No change is observed in the jet flow (Doppler factor) during 2011 in fitted SED. Magnetic field strength in outer zone shows no variation among these states but, SED modeling of inner zone indicates that magnetic field in this zone varies with state. We found comparatively lower magnetic field strength during active states S2 and S4 in inner zone, but found higher magnetic field in high state S5 by SED modeling. We have also not seen any significant change in electron energy distribution of outer zone but change is indicated in inner zone. Higher electron energy density is found at the time of activity in inner zone. Also, we noticed significant change in electron energy spectral index after the break (p2) in this zone.    
 
\begin{table*}
\small
\centering
\caption{{\it Fermi}-LAT spectrum}

\begin{tabular}{ccccc}
\hline
\hline
Epoch & Flux (0.2 - 300 GeV)      & Photon Index & TS & XRT-PCA \\
      & $\times$ $10^{-8}$ ph/cm$^{2}$/sec& 0.2-300 Ge V &    & Index\\
\hline
S1       & 3.8$\pm$0.3   & 1.59 $\pm$ 0.14    & 102   &  1.95     \\
S2       & 5.12$\pm$0.88 & 1.66 $\pm$ 0.13    & 126   &  1.87   \\
S3       & 1.78$\pm$0.34 & 1.27 $\pm$ 0.21    & 67    &  2.11    \\
S4       & 4.5$\pm$0.23  & 1.62 $\pm$ 0.09    & 297   &  1.75 \\
S5       & 5.9$\pm$0.45  & 1.46 $\pm$ 0.10    & 276   &  1.59  \\
\hline
\end{tabular}
\label{FS}
\end{table*}

\section{Discussion}
Mrk~501 is a core dominated radio source, with a one sided jet on pc scale, which extends till $\sim$500 pc. This jet shows several sharp bends followed by rapid expansion and limb brightening structures on physical scale of $\sim$1pc as seen by  \citet{2004ApJ...600..127G}. 

In recent years several attempts were made to model the SED of Mrk~501 with multizone scenario \citep{2005A&A...432..401G,2008ApJ...689...68G,2009MNRAS.395L..29G,2011ApJ...743L..19L} or adding extra break in the injected electron distribution \citep{2011ApJ...727..129A}. Alternatively, the SED could also originate from a two-component spine-sheath structure of the jet transverse to its direction \citep{2005A&A...432..401G}, as suggested by the complex VLBI radio jet morphology of Mrk 501 \citet{2004ApJ...600..127G}. However, the energy transport in blazar jets generally occurs along the jet axis as witnessed by the motion of VLBI radio knots suggesting to logically connect the flux variability with the longitudinal evolution of components. 

In this section, we discuss the implications of the physical parameters of the source resulting from the SSC modeling of the SED. We also try to understand properties of the electron energy distribution emerging from SED modeling and constrain the
physical processes responsible for the particle acceleration. We also examine the broadband variability of Mrk~501 in the framework of our two zone model.

\subsection{Variability}
An alternative way to constrain the physical parameters of the jet is to model its flux variability. Measurement of fast and rapid flux variability can shed light on motions of the bulk outflows of the plasma in innermost region of jets which is well beyond the current imaging capabilities of telescopes in any part of the electromagnetic spectrum.

In the SSC scenario the highest energy tail of the electron energy distribution ($\gamma$$\ge$$\gamma$$_{br}$) is responsible for the production of the observed X-ray synchrotron continuum at $\ge$ 0.5 keV in HBLs, while the TeV $\gamma$-rays might be produced through upscattering of synchrotron photons by the same population of electrons. The observed optical and X-ray variability during 2011-2012 may be explained by injection of fresh electrons in emission zones and cooling of the electrons due to SSC mechanism. Mrk~501 also shows energy dependent flux and spectral variability in $\gamma$-rays. The source flux varies differently from lower energies (0.2-2 GeV) to higher energies (2-300 GeV).  The observed $\gamma$-ray variability is mainly divided into two bands, $<$ 2 GeV and above 2 GeV. The 0.2\,--\,2 GeV $\gamma$-rays observed by {\it Fermi}-LAT could be produced by low-energy electrons through IC scattering of UV synchrotron photons. The observed HE ($>$2 GeV) $\gamma$-rays by {\it Fermi}-LAT and VHE $\gamma$-rays by HAGAR could be produced by IC scattering of the electrons having a Lorentz factor in the range $\sim 10^{4}-10^{5}$.

Mrk~501 was detected in moderate activity state during May 2011 by HAGAR as seen in last panels of Figures \ref{MW_ALL} and \ref{Mw_lc_a}. A positive correlation between low energy X-rays and $\gamma$-rays is seen during this period, with the peak flux being observed in May 2011 in the $\gamma$-ray and X-ray wavebands. Observed flux enhancement during the S2 period can be explained by the injection of fresh electrons in active zone (inner zone) of jet. These electrons may be accelerated to higher energies by shock in the jet. Two zone spectral modeling of this period suggest that the injected electron spectrum in inner zone has a power-law of index 2, which could arise due to Fermi first order mechanism.  Moreover, the {\it Swift}-BAT light curve is found to be anti-correlated with the other wavebands and we can not conclusively say why the flux in {\it Swift}-BAT band is found to be high during S1.

\subsection{Application of two zone model in Mrk~501 multiwaveband data during 2011}

The multifrequency data set provides an opportunity to obtain broad band SED of Mrk~501 in the quiescent and moderate states, and it also allows comparison between these flux states. In the work presented here we have compared five different flux states of Mrk~501 during the year 2011, along with quiescent state SED observed by \citet{2011ApJ...727..129A}. We have modified the single zone model developed by \citet{2004ApJ...601..151K} to a multi zone model and used it to explain broad band emission from Mrk~501. Details of this one zone model can be found in \citet{2004ApJ...601..151K}. Same model was also used by \citet{2012A&A...541A.140S} to explain SED of Mrk~421.  

We have found that $\gamma$-ray emitting zones are very close to the black hole around $\sim$0.08 pc which is consistent with other results presented in literature. The inferred magnetic field from the modeling of SED by our model is also in good agreement with the magnetic field claimed for the partially resolved radio core of Mrk~501 \citep{2004ApJ...600..127G}. But in our model, source is not found in equipartition with the relativistic electrons along with magnetic field. On the other hand, source is consistent with being in equipartition with the relativistic electrons in in spine-sheath model given by \citet{2005A&A...432..401G}. From our study of multiwaveband SED modeling, we also infer that plasma in the jet is moving with Doppler factor of $\sim$ 12 and it does not change with the flux state and time. This value is consistent with other previous works \citep{2011ApJ...727..129A}.  Also we have not detected any change in the speed of the jet in $\gamma$-ray emitting zone ($<$0.1~pc). The electrons in these blobs are accelerated through Fermi first order mechanism producing a power law distribution. We have found the inner blob has much narrower electron distribution than outer blob, with high minimum cutoff for $\gamma_{min}$. 
This blob is responsible for activity in the blazar zone and also the observed hard spectrum in the source. A narrow electron distribution can produce very hard spectrum, similar suggestions were also made by (\cite{2009MNRAS.399L..59T,2006MNRAS.368L..52K,2011ApJ...740...64L}). Electron population of outer zone is evolved and old, and this population has suffered radiative and adiabatic losses. We have observed via SED fitting that contribution of outer zone is not constant and it varies with flux state. Among the six flux states we discussed in this work, we found outer zone contributes significantly to $\gamma$-ray hump in all the states except S5. This could be possible only if outer zone of S5 is not left with high energy electrons at the time of activity. We found that at the time of activity inner blob becomes dominant and it may produce hard spectrum at highest energies. The observed SED of Mrk~501 is sum of total radiation emitted by two zones and the shape of SED depends on the relative contributions from each zone. If both the zones contribute to the total observed flux then SED might be observed with a plateau at lower energy $\gamma$-rays and a hard spectrum at highest energy. The difference between the spectral indices below and above the break energy $\Delta$p=p$_{2}$-p$_{1}$ determined by SED modeling are close to 1 in the case of S1, S2, S3 and MAGIC SED for inner zone, see Table \ref{SED_mrk501}. The value, $\Delta$p=1 is expected as a result of classical synchrotron cooling break for a uniform emission region.

\begin{figure}
\centering
\includegraphics[width=9cm]{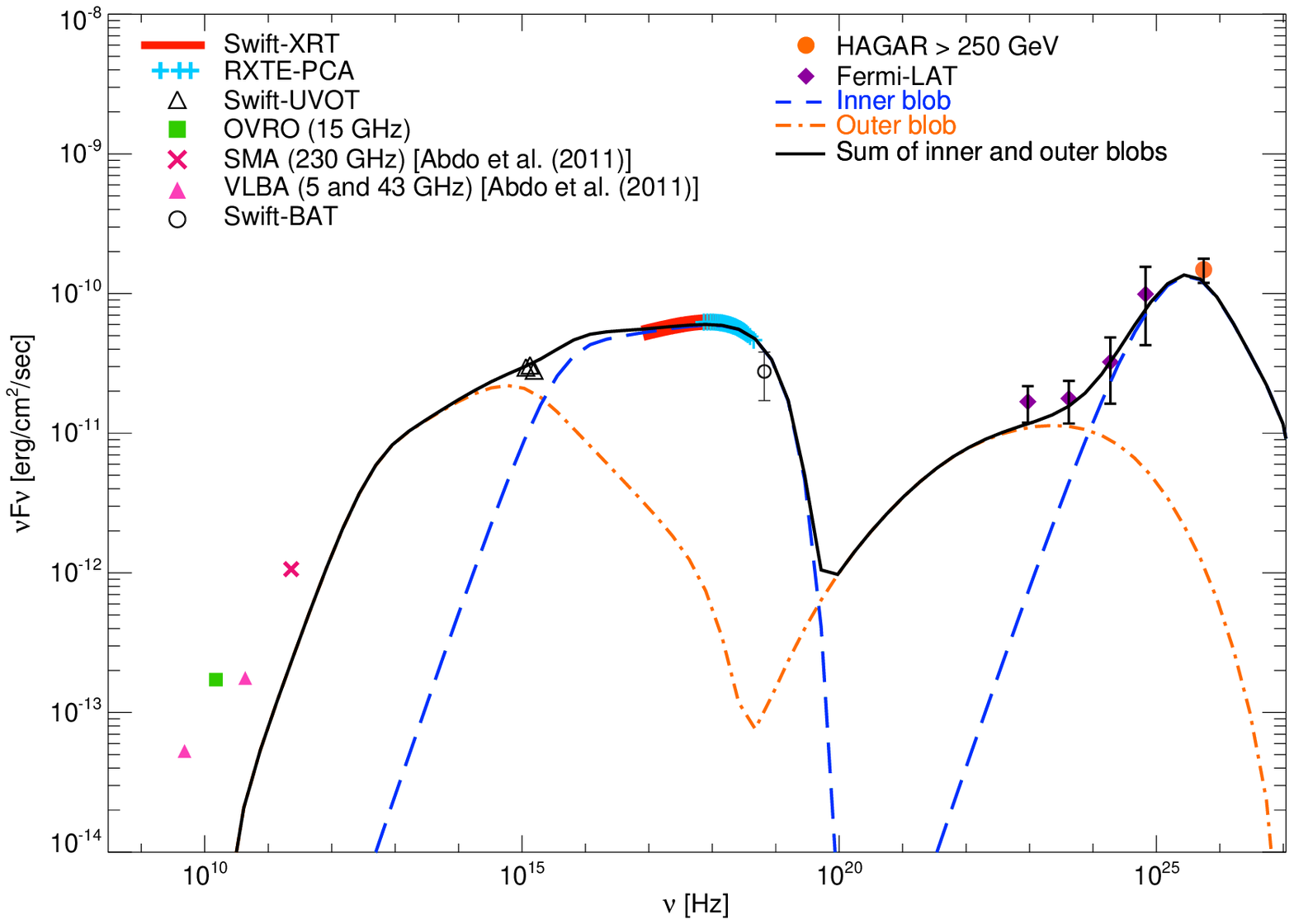}
\caption[]{Two zone SED of Mrk 501 during 2011 April-May [S2]}
\label{SED2}
\end{figure}

The observed LAT spectrum during 2011 May when the source was in moderate bright state, shows plateau in the SED at lower energies $\sim$(0.2-5 GeV), and a break in the slope at $\sim$5 GeV (see Figure \ref{SED2}). A hard spectrum with photon index of 1.5 is detected in higher energy (2-300 GeV) band during these observations. A similar behaviour was also reported during a flare observed in the first 480 days of {\it Fermi}-LAT operation (2008-2009) in \citep{2011ApJ...727..129A}. Spectrum of this flare was found to be very hard/flat ($\sim$1.1) in the 10-200 GeV range. 
The SED in the 0.3-200 GeV range during this flare also shows a break in the slope, around 10 GeV \citep{2012A&A...541A..31N}. Very hard spectra ($<$1.3) are detected in {\it Fermi}-LAT data, few times during 2011-2012 in low energy (0.2-2 GeV) as well as in high energy band (2-300 GeV), (see Figure \ref{Fermi_lat}). Origin of this very hard spectrum is still under debate. 

\begin{deluxetable}{ccccccccccccc}
\tabletypesize{\scriptsize}
\tablecaption{SED parameters \label{SED_mrk501}}
\tablewidth{0pt}
\tablehead{
\colhead{State} & \colhead{T$\_${var}} & \colhead{Magnetic} & \colhead{Doppler} & \colhead{\tablenotemark{1}log E$_{min}$} & \colhead{\tablenotemark{2}log E$_{max}$} & \colhead{\tablenotemark{3}log E$_{break}$}& \colhead{p1} & \colhead{p2} & \colhead{\tablenotemark{4}U$_{e}$} &\colhead{\tablenotemark{5}$\eta$} \\ 
\colhead{-}& \colhead{(hr)}&\colhead{field} & \colhead{factor} & \colhead{[eV]} &\colhead{ [eV]} & \colhead{[eV]} & \colhead{-} &\colhead{ -} & \colhead{[$10^{-3}$]} & \colhead{$[u_{e}^{'}/u_{B}^{'}]$}\\
\colhead{-}&\colhead{-}&\colhead{(G)} & \colhead{($\delta$)} & \colhead{-} & \colhead{-} & \colhead{-}& \colhead{-} & \colhead{-}& \colhead{(erg/cc)} & \colhead{-}  
}
\startdata
\hline
MAGIC  (Outer)&48 & 0.032  & 12.07  & 8.6  & 11.6  & 10.10  & 2.4  & 3.50    &  1.8    &  44.2 \\
S1  (Outer)   &48 & 0.028  & 12.07  & 8.9  & 11.6  & 10.10  & 2.4  & 3.95   &  1.0    &  32.0   \\
S2  (Outer)   &48 & 0.028  & 12.07  & 8.9  & 11.6  & 10.10  & 2.4  & 3.95   &  1.6    &  51.3 \\
S3  (Outer)   &48 &0.028  & 12.07 &  8.9  & 11.6  &  10.10 &  2.4  & 3.95   &  1.0    &  32.0   \\   
S4  (Outer)   &48 & 0.028 & 12.07  & 8.9  & 11.6  & 10.10  & 2.4  & 3.95   &  1.25   &  40.1  \\
S5 (Outer)    &48 & 0.028 & 12.07  & 8.9  & 11.6  & 10.10  & 2.4  & 3.95    & 1.0   &  32.0   \\
\hline
\hline
MAGIC (inner) & 6.9&  0.08& 12.07 & 10.0  & 11.95  & 10.80  &  2.0 & 3.05   & 11   & 43.2\\
S1 (inner)&6.9  & 0.075  & 12.07  & 9.75  & 11.85  & 10.55  & 2.0  & 2.9   & 18   & 80.4\\
S2 (inner)& 6.9 & 0.056  & 12.07  & 10.2 &  11.95 & 10.45  & 2.0  & 2.85   & 27   &  216.0\\
S3 (inner)& 6.9 &  0.075 & 12.07  & 10.2  & 12.00  & 10.45  & 2.0  &  3.1  & 20 & 89.4 \\
S4 (inner)&6.9 & 0.056 & 12.07  & 9.7  & 12.05  & 10.90  & 2.0  & 2.5   &  30   &  240.0 \\
S5 (inner)&6.9 & 0.075 & 12.07  & 8.7  & 11.90  & 11.35  & 2.0  & 2.6   &  32   &  143.0 
\enddata
\tablenotetext{1}{E$_{min}$: Minimum value of energy of the electrons present in the emission zone} \\
\tablenotetext{2}{E$_{max}$: Maximum value of energy of the electron present in the emission zone}  
\tablenotetext{3}{E$_{break}$: Break in the electron injection spectrum} 
\tablenotetext{4}{U$_{e}$: Electron energy density}
\tablenotetext{5}{$\eta$: Equipartition coefficient}
\end{deluxetable}

\begin{figure}
\centering
\includegraphics[width=9cm]{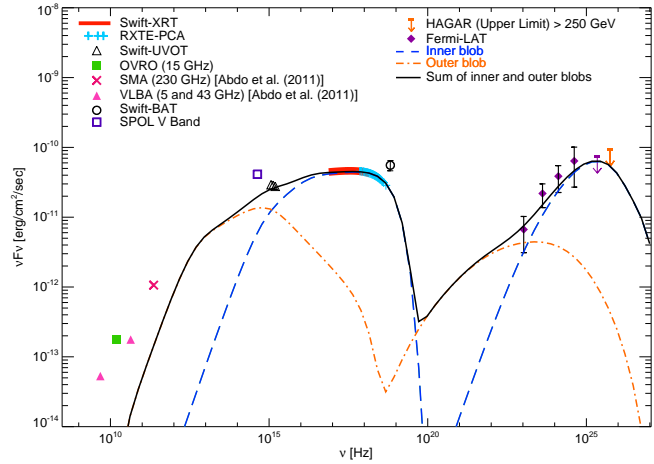}
\caption[]{Two zone SED of Mrk 501 during 2011 March-April [S1]}
\label{SED1}
\end{figure}

\begin{figure}
\centering
\includegraphics[width=9cm]{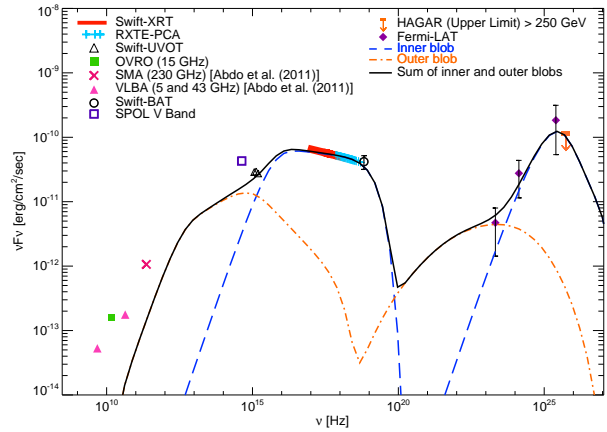}
\caption[]{Two zone SED of Mrk 501 during 2011 June [S3]}
\label{SED3}
\end{figure}

\begin{figure}
\centering
\includegraphics[width=9cm]{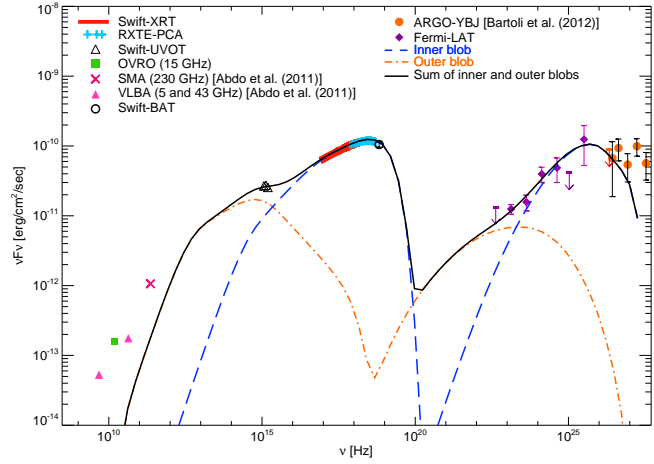}
\caption[]{Two zone SED of Mrk 501 during 2011 November [S4]}
\label{SED_N}
\end{figure}

\acknowledgments
This work used results provided by the ASM/{\it RXTE} teams at MIT. This study also used {\it Swift}/BAT transient monitor results provided by the {\it Swift}/BAT team. This research has also made use of data obtained from the High Energy Astrophysics Science Archive Research Center (HEASARC), provided by NASA’s Goddard Space Flight Center. Data from the Steward Observatory spectropolarimetric monitoring project used supported by {\it Fermi} Guest Investigator grants NNX08AW56G and NNX09AU10G. Radio data at 15 GHz is used from OVRO 40 M Telescope and this {\it Fermi} blazar monitoring program is supported by NASA under award NNX08AW31G, and by the NSF under award 0808050. This research has made use of the XRT Data Analysis Software (XRTDAS) developed under the responsibility of the ASI Science Data Center (ASDC), Italy. We are grateful to the engineering and technical staff of IIA and TIFR, who have taken part in the construction of the HAGAR telescopes and contributed to the setting-up of the front-end electronics and the data acquisition. We also thank Dr. David Paneque and Dr. Chen Songzhan for providing published data of Mrk~501 to use in this study.

\bibliographystyle{apj}
\bibliography{main}

\begin{thebibliography}{60}
	\expandafter\ifx\csname natexlab\endcsname\relax\def\natexlab#1{#1}\fi

	\bibitem[{{Abdo} {et~al.}(2011){Abdo}, {Ackermann}, {Ajello}, {Allafort},
	  {Baldini}, {Ballet}, {Barbiellini}, {Baring}, {Bastieri}, {Bechtol}, \&
	  et~al.}]{2011ApJ...727..129A}
	{Abdo}, A.~A., {et~al.} 2011, \apj, 727, 129

	\bibitem[{{Acciari} {et~al.}(2011){Acciari}, {Arlen}, {Aune}, {Beilicke},
	  {Benbow}, {B{\"o}ttcher}, {Boltuch}, {Bradbury}, {Buckley}, {Bugaev}, \&
	  et~al.}]{2011ApJ...729....2A}
	{Acciari}, V.~A., {et~al.} 2011, \apj, 729, 2

	\bibitem[{{Aharonian} {et~al.}(1999{\natexlab{a}}){Aharonian}, {Akhperjanian},
	  {Barrio}, {Bernl{\"o}hr}, {Bojahr}, {Calle}, {Contreras}, {Cortina}, {Daum},
	  {Deckers}, {Denninghoff}, {Fonseca}, {Gebauer}, {Gonzalez}, {Heinzelmann},
	  {Hemberger}, {Hermann}, {Hess}, {Heusler}, {Hofmann}, {Hohl}, {Horns},
	  {Ibarra}, {Kankanyan}, {Kestel}, {Kirstein}, {K{\"o}hler}, {Kornmayer},
	  {Kranich}, {Krawczynski}, {Lampeitl}, {Lindner}, {Lorenz}, {Magnussen},
	  {Meyer}, {Mirzoyan}, {Moralejo}, {Padilla}, {Panter}, {Petry}, {Plaga},
	  {Plyasheshnikov}, {Prahl}, {P{\"u}hlhofer}, {Rauterberg}, {Renault}, {Rhode},
	  {R{\"o}hring}, {Sahakian}, {Samorski}, {Schmele}, {Schr{\"o}der}, {Stamm},
	  {Wiebel-Sooth}, {Wiedner}, {Willmer}, {Wirth}, \&
	  {Wittek}}]{1999A&A...349...29A}
	{Aharonian}, F., {et~al.} 1999{\natexlab{a}}, \aap, 349, 29

	\bibitem[{{Aharonian} {et~al.}(2001){Aharonian}, {Akhperjanian}, {Barrio},
	  {Bernl{\"o}hr}, {B{\"o}rst}, {Bojahr}, {Bolz}, {Contreras}, {Cortina},
	  {Denninghoff}, {Fonseca}, {Gonzalez}, {G{\"o}tting}, {Heinzelmann},
	  {Hermann}, {Heusler}, {Hofmann}, {Horns}, {Iserlohe}, {Ibarra}, {Jung},
	  {Kankanyan}, {Kestel}, {Kettler}, {Kohnle}, {Konopelko}, {Kornmeyer},
	  {Kranich}, {Krawczynski}, {Lampeitl}, {Lorenz}, {Lucarelli}, {Magnussen},
	  {Mang}, {Meyer}, {Mirzoyan}, {Moralejo}, {Padilla}, {Panter}, {Plaga},
	  {Plyasheshnikov}, {Prahl}, {P{\"u}hlhofer}, {R{\"o}hring}, {Rhode}, {Rowell},
	  {Sahakian}, {Samorski}, {Schilling}, {Schr{\"o}der}, {Siems}, {Stamm},
	  {Tluczykont}, {V{\"o}lk}, {Wiedner}, \& {Wittek}}]{2001ApJ...546..898A}
	---. 2001, \apj, 546, 898

	\bibitem[{{Aharonian}(2000)}]{2000NewA....5..377A}
	{Aharonian}, F.~A. 2000, \nar, 5, 377

	\bibitem[{{Aharonian} {et~al.}(1999{\natexlab{b}}){Aharonian}, {Akhperjanian},
	  {Barrio}, {Bernl{\"o}hr}, {Bojahr}, {Contreras}, {Cortina}, {Daum},
	  {Deckers}, {Fonseca}, {Gonzalez}, {Heinzelmann}, {Hemberger}, {Hermann},
	  {He{\ss}}, {Heusler}, {Hofmann}, {Hohl}, {Horns}, {Ibarra}, {Kankanyan},
	  {Kirstein}, {K{\"o}hler}, {Konopelko}, {Kornmeyer}, {Kranich}, {Krawczynski},
	  {Lampeitl}, {Lindner}, {Lorenz}, {Magnussen}, {Meyer}, {Mirzoyan},
	  {Moralejo}, {Padilla}, {Panter}, {Petry}, {Plaga}, {Plyasheshnikov}, {Prahl},
	  {P{\"u}hlhofer}, {Rauterberg}, {Renault}, {Rhode}, {Sahakian}, {Samorski},
	  {Schmele}, {Schr{\"o}der}, {Stamm}, {V{\"o}lk}, {Wiebel-Sooth}, {Wiedner},
	  {Willmer}, \& {Wirth}}]{1999A&A...342...69A}
	{Aharonian}, F.~A., {et~al.} 1999{\natexlab{b}}, \aap, 342, 69

	\bibitem[{{Albert} {et~al.}(2007){Albert}, {Aliu}, {Anderhub}, {Antoranz},
	  {Armada}, {Baixeras}, {Barrio}, {Bartko}, {Bastieri}, {Becker}, {Bednarek},
	  {Berger}, {Bigongiari}, {Biland}, {Bock}, {Bordas}, {Bosch-Ramon}, {Bretz},
	  {Britvitch}, {Camara}, {Carmona}, {Chilingarian}, {Coarasa}, {Commichau},
	  {Contreras}, {Cortina}, {Costado}, {Curtef}, {Danielyan}, {Dazzi}, {De
	  Angelis}, {Delgado}, {de los Reyes}, {De Lotto}, {Domingo-Santamar{\'{\i}}a},
	  {Dorner}, {Doro}, {Errando}, {Fagiolini}, {Ferenc}, {Fern{\'a}ndez}, {Firpo},
	  {Flix}, {Fonseca}, {Font}, {Fuchs}, {Galante}, {Garc{\'{\i}}a-L{\'o}pez},
	  {Garczarczyk}, {Gaug}, {Giller}, {Goebel}, {Hakobyan}, {Hayashida},
	  {Hengstebeck}, {Herrero}, {H{\"o}hne}, {Hose}, {Hrupec}, {Hsu}, {Jacon},
	  {Jogler}, {Kosyra}, {Kranich}, {Kritzer}, {Laille}, {Lindfors}, {Lombardi},
	  {Longo}, {L{\'o}pez}, {L{\'o}pez}, {Lorenz}, {Majumdar}, {Maneva},
	  {Mannheim}, {Mansutti}, {Mariotti}, {Mart{\'{\i}}nez}, {Mazin}, {Merck},
	  {Meucci}, {Meyer}, {Miranda}, {Mirzoyan}, {Mizobuchi}, {Moralejo}, {Nieto},
	  {Nilsson}, {Ninkovic}, {O{\~n}a-Wilhelmi}, {Otte}, {Oya}, {Paneque},
	  {Panniello}, {Paoletti}, {Paredes}, {Pasanen}, {Pascoli}, {Pauss}, {Pegna},
	  {Persic}, {Peruzzo}, {Piccioli}, {Prandini}, {Puchades}, {Raymers}, {Rhode},
	  {Rib{\'o}}, {Rico}, {Rissi}, {Robert}, {R{\"u}gamer}, {Saggion}, {Saito},
	  {S{\'a}nchez}, {Sartori}, {Scalzotto}, {Scapin}, {Schmitt}, {Schweizer},
	  {Shayduk}, {Shinozaki}, {Shore}, {Sidro}, {Sillanp{\"a}{\"a}}, {Sobczynska},
	  {Stamerra}, {Stark}, {Takalo}, {Tavecchio}, {Temnikov}, {Tescaro}, {Teshima},
	  {Torres}, {Turini}, {Vankov}, {Vitale}, {Wagner}, {Wibig}, {Wittek},
	  {Zandanel}, {Zanin}, \& {Zapatero}}]{2007ApJ...669..862A}
	{Albert}, J., {et~al.} 2007, \apj, 669, 862

	\bibitem[{{Atwood} {et~al.}(2009){Atwood}, {Abdo}, {Ackermann}, {Althouse},
	  {Anderson}, {Axelsson}, {Baldini}, {Ballet}, {Band}, {Barbiellini}, \& et~al.
	  et al. et al. et al. et al. et al. et al. et al. et al. et al. et al. et al.
	  et al. et al. et al. et al. et al. et al. et al. et al. et al. et al. et al.
	  et al. et al. et al. et al. et al. et al. et al. et al. et al. et al. et al.
	  et al. et al. et al. et al. et al. et al. et al. et al. et al. et al. et al.
	  et al. et al. et al. et al. et al. et al. et al. et al. et al. et al. et al.
	  et al. et al. et al. et al. et al. et al. et al.~et
	  al.}]{2009ApJ...697.1071A}
	{Atwood}, W.~B., {et~al.} 2009, \apj, 697, 1071

	\bibitem[{{Bartoli} {et~al.}(2012){Bartoli}, {Bernardini}, {Bi}, {Bleve},
	  {Bolognino}, {Branchini}, {Budano}, {Calabrese Melcarne}, {Camarri}, {Cao},
	  {Cardarelli}, {Catalanotti}, {Cattaneo}, {Chen}, {Chen}, {Chen}, {Creti},
	  {Cui}, {Dai}, {D'Al{\'{\i}} Staiti}, {Danzengluobu}, {Dattoli}, {De Mitri},
	  {D'Ettorre Piazzoli}, {Di Girolamo}, {Ding}, {Di Sciascio}, {Feng}, {Feng},
	  {Feng}, {Galeazzi}, {Giroletti}, {Gou}, {Guo}, {He}, {Hu}, {Hu}, {Huang},
	  {Iacovacci}, {Iuppa}, {James}, {Jia}, {Labaciren}, {Li}, {Li}, {Li},
	  {Liguori}, {Liu}, {Liu}, {Liu}, {Liu}, {Lu}, {Ma}, {Ma}, {Mancarella},
	  {Mari}, {Marsella}, {Martello}, {Mastroianni}, {Montini}, {Ning}, {Pagliaro},
	  {Panareo}, {Panico}, {Perrone}, {Pistilli}, {Ruggieri}, {Salvini},
	  {Santonico}, {Shen}, {Sheng}, {Shi}, {Stanescu}, {Surdo}, {Tan}, {Vallania},
	  {Vernetto}, {Vigorito}, {Wang}, {Wang}, {Wu}, {Wu}, {Xu}, {Xue}, {Yang},
	  {Yang}, {Yao}, {Yuan}, {Zha}, {Zhang}, {Zhang}, {Zhang}, {Zhang}, {Zhang},
	  {Zhang}, {Zhang}, {Zhao}, {Zhaxiciren}, {Zhaxisangzhu}, {Zhou}, {Zhu}, {Zhu},
	  {Zizzi}, \& {ARGO-YBJ Collaboration}}]{2012ApJ...758....2B}
	{Bartoli}, B., {et~al.} 2012, \apj, 758, 2

	\bibitem[{{Begelman} {et~al.}(2008){Begelman}, {Fabian}, \&
	  {Rees}}]{2008MNRAS.384L..19B}
	{Begelman}, M.~C., {Fabian}, A.~C., \& {Rees}, M.~J. 2008, \mnras, 384, L19

	\bibitem[{{Bradt} {et~al.}(1993){Bradt}, {Rothschild}, \&
	  {Swank}}]{1993A&AS...97..355B}
	{Bradt}, H.~V., {Rothschild}, R.~E., \& {Swank}, J.~H. 1993, \aaps, 97, 355

	\bibitem[{{Burrows} {et~al.}(2005){Burrows}, {Hill}, {Nousek}, {Kennea},
	  {Wells}, {Osborne}, {Abbey}, {Beardmore}, {Mukerjee}, {Short}, {Chincarini},
	  {Campana}, {Citterio}, {Moretti}, {Pagani}, {Tagliaferri}, {Giommi},
	  {Capalbi}, {Tamburelli}, {Angelini}, {Cusumano}, {Br{\"a}uninger}, {Burkert},
	  \& {Hartner}}]{2005SSRv..120..165B}
	{Burrows}, D.~N., {et~al.} 2005, \ssr, 120, 165

	\bibitem[{{Cash}(1979)}]{1979ApJ...228..939C}
	{Cash}, W. 1979, \apj, 228, 939

	\bibitem[{{Catanese} {et~al.}(1997){Catanese}, {Bradbury}, {Breslin},
	  {Buckley}, {Carter-Lewis}, {Cawley}, {Dermer}, {Fegan}, {Finley}, {Gaidos},
	  {Hillas}, {Johnson}, {Krennrich}, {Lamb}, {Lessard}, {Macomb}, {McEnery},
	  {Moriarty}, {Quinn}, {Rodgers}, {Rose}, {Samuelson}, {Sembroski},
	  {Srinivasan}, {Weekes}, \& {Zweerink}}]{1997ApJ...487L.143C}
	{Catanese}, M., {et~al.} 1997, \apjl, 487, L143

	\bibitem[{{Cerruti} {et~al.}(2012){Cerruti}, {Zech}, {Boisson}, \&
	  {Inoue}}]{2012AIPC.1505..635C}
	{Cerruti}, M., {Zech}, A., {Boisson}, C., \& {Inoue}, S. 2012, in American
	  Institute of Physics Conference Series, Vol. 1505, American Institute of
	  Physics Conference Series, ed. F.~A. {Aharonian}, W.~{Hofmann}, \& F.~M.
	  {Rieger}, 635--638

	\bibitem[{{Chitnis} {et~al.}(2009){Chitnis}, {Pendharkar}, {Bose}, {Agrawal},
	  {Rao}, \& {Misra}}]{2009ApJ...698.1207C}
	{Chitnis}, V.~R., {Pendharkar}, J.~K., {Bose}, D., {Agrawal}, V.~K., {Rao},
	  A.~R., \& {Misra}, R. 2009, \apj, 698, 1207

	\bibitem[{{Crusius-Waetzel} \& {Lesch}(1998)}]{1998A&A...338..399C}
	{Crusius-Waetzel}, A.~R., \& {Lesch}, H. 1998, \aap, 338, 399

	\bibitem[{{Dermer} \& {Schlickeiser}(1993)}]{1993ApJ...416..458D}
	{Dermer}, C.~D., \& {Schlickeiser}, R. 1993, \apj, 416, 458

	\bibitem[{{Georganopoulos} \& {Kazanas}(2003)}]{2003ApJ...594L..27G}
	{Georganopoulos}, M., \& {Kazanas}, D. 2003, \apjl, 594, L27

	\bibitem[{{Ghisellini} {et~al.}(2002){Ghisellini}, {Celotti}, \&
	  {Costamante}}]{2002A&A...386..833G}
	{Ghisellini}, G., {Celotti}, A., \& {Costamante}, L. 2002, \aap, 386, 833

	\bibitem[{{Ghisellini} \& {Madau}(1996)}]{1996MNRAS.280...67G}
	{Ghisellini}, G., \& {Madau}, P. 1996, \mnras, 280, 67

	\bibitem[{{Ghisellini} {et~al.}(2009){Ghisellini}, {Tavecchio}, {Bodo}, \&
	  {Celotti}}]{2009MNRAS.393L..16G}
	{Ghisellini}, G., {Tavecchio}, F., {Bodo}, G., \& {Celotti}, A. 2009, \mnras,
	  393, L16

	\bibitem[{{Ghisellini} {et~al.}(2005){Ghisellini}, {Tavecchio}, \&
	  {Chiaberge}}]{2005A&A...432..401G}
	{Ghisellini}, G., {Tavecchio}, F., \& {Chiaberge}, M. 2005, \aap, 432, 401

	\bibitem[{{Giannios} {et~al.}(2009){Giannios}, {Uzdensky}, \&
	  {Begelman}}]{2009MNRAS.395L..29G}
	{Giannios}, D., {Uzdensky}, D.~A., \& {Begelman}, M.~C. 2009, \mnras, 395, L29

	\bibitem[{{Giannios} {et~al.}(2010){Giannios}, {Uzdensky}, \&
	  {Begelman}}]{2010MNRAS.402.1649G}
	---. 2010, \mnras, 402, 1649

	\bibitem[{{Giroletti} {et~al.}(2004){Giroletti}, {Giovannini}, {Feretti},
	  {Cotton}, {Edwards}, {Lara}, {Marscher}, {Mattox}, {Piner}, \&
	  {Venturi}}]{2004ApJ...600..127G}
	{Giroletti}, M., {et~al.} 2004, \apj, 600, 127

	\bibitem[{{Gliozzi} {et~al.}(2006){Gliozzi}, {Sambruna}, {Jung}, {Krawczynski},
	  {Horan}, \& {Tavecchio}}]{2006ApJ...646...61G}
	{Gliozzi}, M., {Sambruna}, R.~M., {Jung}, I., {Krawczynski}, H., {Horan}, D.,
	  \& {Tavecchio}, F. 2006, \apj, 646, 61

	\bibitem[{{Gothe} {et~al.}(2013){Gothe}, {Prabhu}, {Vishwanath}, {Acharya},
	  {Srinivasan}, {Chitnis}, {Kamath}, {Srinivasulu}, {Saleem}, {Kemkar},
	  {Mahesh}, {Gabriel}, {Manoharan}, {Dorji}, {Dorjai}, {Angchuk}, {D'souza},
	  {Duhan}, {Nagesh}, {Rao}, {Sharma}, {Singh}, {Sudersanan}, {Thsering},
	  {Upadhya}, {Anupama}, {Britto}, {Cowsik}, {Saha}, \&
	  {Shukla}}]{2013ExA....35..489G}
	{Gothe}, K.~S., {et~al.} 2013, Experimental Astronomy, 35, 489

	\bibitem[{{Graff} {et~al.}(2008){Graff}, {Georganopoulos}, {Perlman}, \&
	  {Kazanas}}]{2008ApJ...689...68G}
	{Graff}, P.~B., {Georganopoulos}, M., {Perlman}, E.~S., \& {Kazanas}, D. 2008,
	  \apj, 689, 68

	\bibitem[{{Gupta} {et~al.}(2008){Gupta}, {Deng}, {Joshi}, {Bai}, \&
	  {Lee}}]{2008NewA...13..375G}
	{Gupta}, A.~C., {Deng}, W.~G., {Joshi}, U.~C., {Bai}, J.~M., \& {Lee}, M.~G.
	  2008, \nar, 13, 375

	\bibitem[{{Gupta} {et~al.}(2012){Gupta}, {Pandey}, {Singh}, {Rani}, {Pan},
	  {Fan}, \& {Gupta}}]{2012NewA...17....8G}
	{Gupta}, S.~P., {Pandey}, U.~S., {Singh}, K., {Rani}, B., {Pan}, J., {Fan},
	  J.~H., \& {Gupta}, A.~C. 2012, \nar, 17, 8

	\bibitem[{{Kalberla} {et~al.}(2005){Kalberla}, {Burton}, {Hartmann}, {Arnal},
	  {Bajaja}, {Morras}, \& {P{\"o}ppel}}]{2005A&A...440..775K}
	{Kalberla}, P.~M.~W., {Burton}, W.~B., {Hartmann}, D., {Arnal}, E.~M.,
	  {Bajaja}, E., {Morras}, R., \& {P{\"o}ppel}, W.~G.~L. 2005, \aap, 440, 775

	\bibitem[{{Katarzy{\'n}ski} {et~al.}(2006){Katarzy{\'n}ski}, {Ghisellini},
	  {Tavecchio}, {Gracia}, \& {Maraschi}}]{2006MNRAS.368L..52K}
	{Katarzy{\'n}ski}, K., {Ghisellini}, G., {Tavecchio}, F., {Gracia}, J., \&
	  {Maraschi}, L. 2006, \mnras, 368, L52

	\bibitem[{{Katarzy{\'n}ski} {et~al.}(2001){Katarzy{\'n}ski}, {Sol}, \&
	  {Kus}}]{2001A&A...367..809K}
	{Katarzy{\'n}ski}, K., {Sol}, H., \& {Kus}, A. 2001, \aap, 367, 809

	\bibitem[{{Krawczynski}(2004)}]{2004NewAR..48..367K}
	{Krawczynski}, H. 2004, \nar, 48, 367

	\bibitem[{{Krawczynski} {et~al.}(2000){Krawczynski}, {Coppi}, {Maccarone}, \&
	  {Aharonian}}]{2000A&A...353...97K}
	{Krawczynski}, H., {Coppi}, P.~S., {Maccarone}, T., \& {Aharonian}, F.~A. 2000,
	  \aap, 353, 97

	\bibitem[{{Krawczynski} {et~al.}(2004){Krawczynski}, {Hughes}, {Horan},
	  {Aharonian}, {Aller}, {Aller}, {Boltwood}, {Buckley}, {Coppi}, {Fossati},
	  {G{\"o}tting}, {Holder}, {Horns}, {Kurtanidze}, {Marscher}, {Nikolashvili},
	  {Remillard}, {Sadun}, \& {Schr{\"o}der}}]{2004ApJ...601..151K}
	{Krawczynski}, H., {et~al.} 2004, \apj, 601, 151

	\bibitem[{{Krimm} {et~al.}(2013){Krimm}, {Holland}, {Corbet}, {Pearlman},
	  {Romano}, {Kennea}, {Bloom}, {Barthelmy}, {Baumgartner}, {Cummings},
	  {Gehrels}, {Lien}, {Markwardt}, {Palmer}, {Sakamoto}, {Stamatikos}, \&
	  {Ukwatta}}]{2013ApJS..209...14K}
	{Krimm}, H.~A., {et~al.} 2013, \apjs, 209, 14

	\bibitem[{{Krishan} \& {Wiita}(1994)}]{1994ApJ...423..172K}
	{Krishan}, V., \& {Wiita}, P.~J. 1994, \apj, 423, 172

	\bibitem[{{Lefa} {et~al.}(2011{\natexlab{a}}){Lefa}, {Aharonian}, \&
	  {Rieger}}]{2011ApJ...743L..19L}
	{Lefa}, E., {Aharonian}, F.~A., \& {Rieger}, F.~M. 2011{\natexlab{a}}, \apjl,
	  743, L19

	\bibitem[{{Lefa} {et~al.}(2011{\natexlab{b}}){Lefa}, {Rieger}, \&
	  {Aharonian}}]{2011ApJ...740...64L}
	{Lefa}, E., {Rieger}, F.~M., \& {Aharonian}, F. 2011{\natexlab{b}}, \apj, 740,
	  64

	\bibitem[{{Levinson}(2007)}]{2007ApJ...671L..29L}
	{Levinson}, A. 2007, \apjl, 671, L29

	\bibitem[{{Mannheim}(1998)}]{1998Sci...279..684M}
	{Mannheim}, K. 1998, Science, 279, 684

	\bibitem[{{Mattox} {et~al.}(1996){Mattox}, {Bertsch}, {Chiang}, {Dingus},
	  {Digel}, {Esposito}, {Fierro}, {Hartman}, {Hunter}, {Kanbach}, {Kniffen},
	  {Lin}, {Macomb}, {Mayer-Hasselwander}, {Michelson}, {von Montigny},
	  {Mukherjee}, {Nolan}, {Ramanamurthy}, {Schneid}, {Sreekumar}, {Thompson}, \&
	  {Willis}}]{1996ApJ...461..396M}
	{Mattox}, J.~R., {et~al.} 1996, \apj, 461, 396

	\bibitem[{{M{\"u}cke} {et~al.}(2003){M{\"u}cke}, {Protheroe}, {Engel},
	  {Rachen}, \& {Stanev}}]{2003APh....18..593M}
	{M{\"u}cke}, A., {Protheroe}, R.~J., {Engel}, R., {Rachen}, J.~P., \& {Stanev},
	  T. 2003, Astroparticle Physics, 18, 593

	\bibitem[{{Neronov} {et~al.}(2012){Neronov}, {Semikoz}, \&
	  {Taylor}}]{2012A&A...541A..31N}
	{Neronov}, A., {Semikoz}, D., \& {Taylor}, A.~M. 2012, \aap, 541, A31

	\bibitem[{{Pian} {et~al.}(1998){Pian}, {Vacanti}, {Tagliaferri}, {Ghisellini},
	  {Maraschi}, {Treves}, {Urry}, {Fiore}, {Giommi}, {Palazzi}, {Chiappetti}, \&
	  {Sambruna}}]{1998ApJ...492L..17P}
	{Pian}, E., {et~al.} 1998, \apjl, 492, L17

	\bibitem[{{Quinn} {et~al.}(1996){Quinn}, {Akerlof}, {Biller}, {Buckley},
	  {Carter-Lewis}, {Cawley}, {Catanese}, {Connaughton}, {Fegan}, {Finley},
	  {Gaidos}, {Hillas}, {Lamb}, {Krennrich}, {Lessard}, {McEnery}, {Meyer},
	  {Mohanty}, {Rodgers}, {Rose}, {Sembroski}, {Schubnell}, {Weekes}, {Wilson},
	  \& {Zweerink}}]{1996ApJ...456L..83Q}
	{Quinn}, J., {et~al.} 1996, \apjl, 456, L83

	\bibitem[{{Richards} {et~al.}(2011){Richards}, {Max-Moerbeck}, {Pavlidou},
	  {King}, {Pearson}, {Readhead}, {Reeves}, {Shepherd}, {Stevenson},
	  {Weintraub}, {Fuhrmann}, {Angelakis}, {Zensus}, {Healey}, {Romani}, {Shaw},
	  {Grainge}, {Birkinshaw}, {Lancaster}, {Worrall}, {Taylor}, {Cotter}, \&
	  {Bustos}}]{2011ApJS..194...29R}
	{Richards}, J.~L., {et~al.} 2011, \apjs, 194, 29

	\bibitem[{{Roming} {et~al.}(2005){Roming}, {Kennedy}, {Mason}, {Nousek}, {Ahr},
	  {Bingham}, {Broos}, {Carter}, {Hancock}, {Huckle}, {Hunsberger}, {Kawakami},
	  {Killough}, {Koch}, {McLelland}, {Smith}, {Smith}, {Soto}, {Boyd},
	  {Breeveld}, {Holland}, {Ivanushkina}, {Pryzby}, {Still}, \&
	  {Stock}}]{2005SSRv..120...95R}
	{Roming}, P.~W.~A., {et~al.} 2005, \ssr, 120, 95

	\bibitem[{{Saha} {et~al.}(2013){Saha}, {Chitnis}, {Vishwanath}, {Kale},
	  {Shukla}, {Acharya}, {Anupama}, {Bhattacharjee}, {Britto}, {Prabhu}, \&
	  {Singh}}]{2013APh....42...33S}
	{Saha}, L., {et~al.} 2013, Astroparticle Physics, 42, 33

	\bibitem[{{Sambruna} {et~al.}(2000){Sambruna}, {Aharonian}, {Krawczynski},
	  {Akhperjanian}, {Barrio}, {Bernl{\"o}hr}, {Bojahr}, {Calle}, {Contreras},
	  {Cortina}, {Denninghoff}, {Fonseca}, {Gonzalez}, {G{\"o}tting},
	  {Heinzelmann}, {Hemberger}, {Hermann}, {Heusler}, {Hofmann}, {Horns},
	  {Ibarra}, {Kankanyan}, {Kestel}, {Kettler}, {K{\"o}hler}, {Kohnle},
	  {Konopelko}, {Kornmeyer}, {Kranich}, {Lampeitl}, {Lindner}, {Lorenz},
	  {Magnussen}, {Mang}, {Meyer}, {Mirzoyan}, {Moralejo}, {Padilla}, {Panter},
	  {Plaga}, {Plyasheshnikov}, {Prahl}, {P{\"u}hlhofer}, {Rauterberg},
	  {R{\"o}hring}, {Sahakian}, {Samorski}, {Schilling}, {Schmele},
	  {Schr{\"o}der}, {Stamm}, {Tluczykont}, {V{\"o}lk}, {Wiebel-Sooth}, {Wiedner},
	  {Willmer}, {Wittek}, {Chou}, {Coppi}, {Rothschild}, \&
	  {Urry}}]{2000ApJ...538..127S}
	{Sambruna}, R.~M., {et~al.} 2000, \apj, 538, 127

	\bibitem[{{Schlegel} {et~al.}(1998){Schlegel}, {Finkbeiner}, \&
	  {Davis}}]{1998ApJ...500..525S}
	{Schlegel}, D.~J., {Finkbeiner}, D.~P., \& {Davis}, M. 1998, \apj, 500, 525

	\bibitem[{{Shukla} {et~al.}(2012){Shukla}, {Chitnis}, {Vishwanath}, {Acharya},
	  {Anupama}, {Bhattacharjee}, {Britto}, {Prabhu}, {Saha}, \&
	  {Singh}}]{2012A&A...541A.140S}
	{Shukla}, A., {et~al.} 2012, \aap, 541, A140

	\bibitem[{{Smith} {et~al.}(2009){Smith}, {Montiel}, {Rightley}, {Turner},
	  {Schmidt}, \& {Jannuzi}}]{2009arXiv0912.3621S}
	{Smith}, P.~S., {Montiel}, E., {Rightley}, S., {Turner}, J., {Schmidt}, G.~D.,
	  \& {Jannuzi}, B.~T. 2009, ArXiv e-prints

	\bibitem[{{Subramanian} {et~al.}(2012){Subramanian}, {Shukla}, \&
	  {Becker}}]{2012MNRAS.423.1707S}
	{Subramanian}, P., {Shukla}, A., \& {Becker}, P.~A. 2012, \mnras, 423, 1707

	\bibitem[{{Tavecchio} {et~al.}(2009){Tavecchio}, {Ghisellini}, {Ghirlanda},
	  {Costamante}, \& {Franceschini}}]{2009MNRAS.399L..59T}
	{Tavecchio}, F., {Ghisellini}, G., {Ghirlanda}, G., {Costamante}, L., \&
	  {Franceschini}, A. 2009, \mnras, 399, L59

	\bibitem[{{Tavecchio} {et~al.}(2001){Tavecchio}, {Maraschi}, {Pian},
	  {Chiappetti}, {Celotti}, {Fossati}, {Ghisellini}, {Palazzi}, {Raiteri},
	  {Sambruna}, {Treves}, {Urry}, {Villata}, \&
	  {Djannati-Ata{\"i}}}]{2001ApJ...554..725T}
	{Tavecchio}, F., {et~al.} 2001, \apj, 554, 725

	\bibitem[{{Villata} \& {Raiteri}(1999)}]{1999A&A...347...30V}
	{Villata}, M., \& {Raiteri}, C.~M. 1999, \aap, 347, 30

	\bibitem[{{Xue} \& {Cui}(2005)}]{2005ApJ...622..160X}
	{Xue}, Y., \& {Cui}, W. 2005, \apj, 622, 160

	\end{thebibliography}
\end{document}